\documentclass[twocolumn]{aastex631}

\usepackage{blindtext}
\usepackage{longtable}
\usepackage{graphicx}
\usepackage{amssymb}
\usepackage{amsmath}
\usepackage{rotating}
\usepackage{pifont}
\newcommand{\cmark}{\ding{51}}%
\newcommand{\xmark}{\ding{55}}%

\def\lya{Ly$\alpha$}

\makeatletter
\DeclareRobustCommand{\NHI}{%
  $N_{\mbox{\scriptsize H\,\check@mathfonts\fontsize\sf@size\z@\selectfont I}}$}
\makeatother

\makeatletter
\DeclareRobustCommand{\OmHI}{%
  $\Omega_{\mbox{\scriptsize H\,\check@mathfonts\fontsize\sf@size\z@\selectfont I}}$}
\makeatother

\makeatletter
\DeclareRobustCommand{\OmHIa}{%
  $\Omega^{\mathcal{A}}_{\mbox{\scriptsize H\,\check@mathfonts\fontsize\sf@size\z@\selectfont I}}$}
\makeatother

\makeatletter
\DeclareRobustCommand{\OmHIb}{%
  $\Omega^{\mathcal{B}}_{\mbox{\scriptsize H\,\check@mathfonts\fontsize\sf@size\z@\selectfont I}}$}
\makeatother

\makeatletter
\DeclareRobustCommand{\OmHIc}{%
  $\Omega^{\mathcal{C}}_{\mbox{\scriptsize H\,\check@mathfonts\fontsize\sf@size\z@\selectfont I}}$}
\makeatother

\makeatletter
\DeclareRobustCommand{\OmHId}{%
  $\Omega^{\mathcal{D}}_{\mbox{\scriptsize H\,\check@mathfonts\fontsize\sf@size\z@\selectfont I}}$}
\makeatother

\makeatletter
\DeclareRobustCommand{\HI}{%
  \mbox{H\,\check@mathfonts\fontsize\sf@size\z@\selectfont I}%
}
\makeatother

\makeatletter
\DeclareRobustCommand{\MgII}{%
  \mbox{Mg\,\check@mathfonts\fontsize\sf@size\z@\selectfont II}%
}
\makeatother

\makeatletter
\DeclareRobustCommand{\HII}{%
  \mbox{H\,\check@mathfonts\fontsize\sf@size\z@\selectfont II}%
}
\makeatother

\def\SFRD{$\psi{_*}$}
\def\rhoHI{$\rho_{\mbox{\tiny \HI}}$}
\def\rhoHIa{$\rho_{\mbox{\tiny \HI}}^{\mathcal{A}}$}
\def\rhoHIb{$\rho_{\mbox{\tiny \HI}}^{\mathcal{B}}$}
\received{October 14th, 2024}
\revised{January 23rd, 2025}
\accepted{February 6th, 2025}

\shorttitle{The Qz5 Survey (I): How the HI Mass Density of the Universe Evolves With Cosmic Time}

\shortauthors{Oyarz\'un et al.}

\begin{document}

\title{The Qz5 Survey (I): How the HI Mass Density of the Universe Evolves With Cosmic Time}

\email{goyarzu1@jhu.edu}

\author[0000-0003-0028-4130]{Grecco A. Oyarz\'un}
\affiliation{Department of Physics and Astronomy, Johns Hopkins University, Baltimore, MD 21218, USA}

\author[0000-0002-9946-4731]{Marc Rafelski}
\affiliation{Space Telescope Science Institute, 3700 San Martin Drive, Baltimore, MD 21218, USA}
\affiliation{Department of Physics and Astronomy, Johns Hopkins University, Baltimore, MD 21218, USA}

\author[0000-0001-8415-7547]{Lise Christensen}
\affiliation{Cosmic Dawn Center, Niels Bohr Institute, University of Copenhagen, Jagtvej 128, 2200-N Copenhagen, Denmark}

\author[0000-0002-1155-977X]{Fiona Ozyurt}
\affiliation{Maria Mitchell Observatory, 4 Vestal St. Nantucket, MA 02554, USA}
\affiliation{Department of Physics, Southern Connecticut State University, New Haven, CT 06515, USA}

\author[0000-0003-2973-0472]{Regina A. Jorgenson}
\affiliation{Maria Mitchell Observatory, 4 Vestal St. Nantucket, MA 02554, USA}

\author[0000-0002-9838-8191]{M. Neeleman}
\affiliation{National Radio Astronomy Observatory, 520 Edgemont Road, Charlottesville, VA, 22903, USA}

\author[0000-0001-6676-3842]{Michele Fumagalli}
\affiliation{Dipartimento di Fisica G. Occhialini, Universit\`a degli Studi di Milano-Bicocca, Piazza della Scienza 3, I-20126 Milano, Italy}
\affiliation{INAF - Osservatorio Astronomico di Trieste, via G. B. Tiepolo 11, I-34143 Trieste, Italy}

\author[0000-0002-7738-6875]{J. Xavier Prochaska}
\affiliation{Department of Astronomy \& Astrophysics, UCO/Lick Observatory, University of California, 1156 High Street, Santa Cruz, CA 95064, USA}
\affiliation{Kavli Institute for the Physics and Mathematics of the Universe (Kavli IPMU), 5-1-5 Kashiwanoha, Kashiwa, 277-8583, Japan}
\affiliation{Division of Science, National Astronomical Observatory of Japan, 2-21-1 Osawa, Mitaka, Tokyo 181-8588, Japan}

\author[0000-0003-0960-3580]{G. Worseck}
\affiliation{Institut für Physik und Astronomie, Universität Potsdam, Karl-Liebknecht-Str. 24/25, D-14476 Potsdam, Germany}

\author[0000-0002-6505-9981]{M.E. Wisz}
\affiliation{Maria Mitchell Observatory, 4 Vestal St. Nantucket, MA 02554, USA}
\affiliation{1Department of Physics, University of California, Merced, CA 95343, USA}

\author[0000-0003-2344-263X]{George D. Becker} 
\affiliation{Department of Physics \& Astronomy, University of California, Riverside, CA, 92521, USA}

\author[0000-0003-0389-0902]{Sebasti\'an L\'opez} 
\affiliation{Departamento de Astronom\'ia, Universidad de Chile, Casilla 36-D, Santiago, Chile}



\begin{abstract}
We report that the neutral hydrogen (\HI) mass density of the Universe (\rhoHI) increases with cosmic time since $z\sim 5$, peaks at $z\sim 3$, and then decreases toward $z\sim 0$. This is the first result of Qz5, our spectroscopic survey of 63 quasars at $z\gtrsim 5$ with \mbox{VLT/X-SHOOTER} and Keck/ESI aimed at characterizing intervening \HI\ gas absorbers at $z\sim 5$. The main feature of Qz5 is the high resolution ($R\sim7000-9000$) of the spectra, which allows us to (1) accurately detect high column density \HI\ gas absorbers in an increasingly neutral intergalactic medium at $z\sim 5$ and (2) determine the reliability of previous \rhoHI\ measurements derived with lower resolution spectroscopy. We find 5 intervening Damped \lya\ absorbers (DLAs) at $z>4.5$, which corresponds to the lowest DLA incidence rate ($0.034_{0.02}^{0.05}$) at $z\gtrsim 2$. We also measure the lowest \rhoHI\ at $z\gtrsim 2$ from our sample of DLAs and subDLAs, corresponding to \rhoHI~$=0.56^{0.82}_{0.31}\times 10^8~\mbox{M}_{\odot}$~Mpc$^{-3}$ at $z\sim 5$. Taking into account our measurements at $z\sim 5$ and systematic biases in the DLA detection rate at lower spectral resolutions, we conclude that \rhoHI\ doubles from $z\sim 5$ to $z\sim 3$. From these results emerges a qualitative agreement between how the cosmic densities of \HI\ gas mass, molecular gas mass, and star-formation rate build up with cosmic time.
\end{abstract}


\section{Introduction} 
\label{1}

The gas within and around galaxies is a critical component in our models of galaxy formation and evolution. The exchange of gas between galaxies and their circumgalactic medium (CGM) --- i.e., the baryon cycle --- is necessary to explain how galaxies modulate their star-formation, how they transition into quiescence, and how their chemical composition (both stellar and gaseous) evolves with cosmic time (\citealt{tumlinson2017}). Particularly responsible for triggering star-formation is neutral hydrogen (\HI). Because of its low temperatures ($T<10^4~\mathrm{K}$), \HI\ can be efficiently accreted onto galaxies and ultimately form molecular gas (H$_2$) where high enough densities for star formation are achieved (\citealt{krumholz2011,krumholz2012,glover-clark2012}). These processes are imprinted in observations of nearby galaxies, with star-forming galaxies often showing evidence of high \HI\ mass surface densities (e.g., \citealt{kennicutt1998,kennicutt2012,fraternali2002,sancisi2008}).

Key for the formation of this picture have been observations of gas in emission (\citealt{haynes1984,zwaan2005,bigiel2008,cantinella2013,cantinella2018,haynes2018}). The \HI\ emission line at 21~cm has been exploited with great success to estimate the star-formation efficiency, size, and kinematics of \HI\ disks in the nearby Universe \citep[e.g.][]{leroy2008,wang2016b,walter2008,begum2008,niankun2022,sharma2023,dou2024}. Unfortunately, the 21~cm line is too weak to directly measure with current observatories in individual galaxies beyond $z>0.4$ \citep{fernandez2016,xi2024} and through stacking beyond $z\sim1.5$ \citep{chowdhury2020,chowdhury2021,chowdhury2022}. Thus, we remain unable to constrain the properties of \HI\ through the peak of cosmic star-formation history in emission ($z\gtrsim 2$; \citealt{madau-dickinson2014}).

To study \HI\ at $z= 2-5$, we can instead utilize observations of \HI\ in absorption. The principle behind \HI\ absorption studies is that neutral gas throughout the Universe absorbs and/or scatters most of the UV radiation field. Primary examples of this technique are the absorption signatures produced by high column density \HI\ gas in the spectra of background quasars (QSO), known as a Damped Ly$\alpha$ Absorbers (DLAs; \citealt{wolfe2005}). DLAs provide a unique laboratory for studying \HI\ gas at intermediate and high redshifts, enabling measurements of the dust depletion, molecular fraction, gas temperature, gas kinematics, and metal enrichment of \HI\ gas reservoirs out to $z\sim 5.5$ (e.g. \citealt{prochaska-wolfe1997,rafelski2012,rafelski2014,prochaska2013,neeleman2013,neeleman2015,crighton2015,noterdaeme2015,noterdaeme2023,balashev2017,bolmer2019,combes2024}). Moreover, DLAs are also regularly exploited to search for the galaxies associated with \HI\ gas reservoirs out to $z\sim 4.5$ (i.e., \HI-selected galaxies; e.g. \citealt{moller-warren1993,chen2003,fynbo2010,krogager2017,neeleman2017,neeleman2019,mackenzie2019,kanekar2020,kaur2022b,lofthouse2023,oyarzun2024}).

Because of their high \NHI, DLAs are thought to account for the bulk ($>70\%$) of the \HI\ mass from $z = 2$ (e.g. \citealt{omeara2007,zafar2013}) out to at least $z\sim 4$ (\citealt{berg2019}). This makes the DLAs a convenient probe for the cosmic \HI\ mass density (\rhoHI). This quantity garners interest from both observational and theoretical work (e.g. \citealt{prochaska-wolfe2009,noterdaeme2009,noterdaeme2012,rahmati2015,popping2014}) because it captures the availability of fuel for star-formation throughout cosmic time and because it reflects the efficiency with which gas is turned into stars in the Universe (e.g. \citealt{walter2020}).

While semi-analytic models and some simulations predict a rise and fall of \rhoHI\ similar to the cosmic star-formation rate density (e.g. \citealt{Somerville2001, popping2014, dave2019}), the interpretation of most DLA surveys to date is that \rhoHI\ is a steadily increasing function with redshift out to at least $z\sim 5$ (e.g., \citealt{prochaska-wolfe2009,noterdaeme2012,crighton2015}), with the functional form $\mbox{\rhoHI}\propto (1+z)^{0.57}$ providing a reasonable description of the observations (\citealt{peroux-howk2020}). 

To better understand the tension between simulations and observations in how \rhoHI\ evolves with redshift, we need to establish if there are any systematic biases affecting the different techniques employed for measuring \rhoHI. Relevant to this point is that attaining large, statistically significant samples of quasars (QSOs) at $z\gtrsim 5$ is difficult due to the decrease in both the spatial density and brightness of QSOs with redshift. It is because of these challenges that DLA surveys designed to measure \rhoHI\ at $z\gtrsim 3.5$ often turn to low resolution ($R\sim2000$) and low signal-to-noise (S/N~$\lesssim 5$) spectroscopy (\citealt{prochaska-wolfe2009,bird2017,ho-bird-garnett2020}). Even at higher signal-to-noise (S/N~$\sim 15$) and after careful simulations to account for false positives and incompleteness at low resolutions, \citet{crighton2015} still found the DLA incidence rate to remain highly uncertain for $R\sim1300$ data. This is because the intergalactic medium (IGM) becomes increasingly opaque as redshift increases, making spectroscopic identification of DLAs toward $z\gtrsim 3.5$ more challenging than at $z\sim 2-3$ (\citealt{rafelski2012,noterdaeme2012}). Therefore, our understanding of how \rhoHI\ evolves with redshift would greatly benefit from a high resolution spectroscopic survey of QSOs at $z\gtrsim 3.5$.

In this paper, we present Qz5\footnote{\url{https://doi.org/10.5281/zenodo.14825981}}: a high-resolution ($R\sim7000-9000$) spectroscopic survey of 63 bright QSOs at $z\gtrsim 5$ specifically designed to measure (1) \rhoHI\ at $z\sim 5$ and (2) the metallicity of the Universe at $z\sim 5$ (Wisz et al.~in prep). The high-resolution spectroscopic observations were carried with either Keck/ESI (\citealt{Keck-ESI}) or VLT/X-SHOOTER (\citealt{x-shooter}). The paper is structured as follows. In Section \ref{2} we introduce our sample and describe the observations. Our methodology to identify DLAs and to measure \rhoHI\ is presented in Section \ref{3}. Our results are outlined in Section \ref{4}. We conclude in Section \ref{5} and summarize in Section \ref{6}. We adopt a $\Lambda$CDM cosmology with $\Omega_{\Lambda} = 0.7$, $\Omega_{m} = 0.3$, and $H_0 = 70~$km~s$^{-1}$~Mpc$^{-1}$. All cosmic mass densities are reported in comoving units and all magnitudes are reported in the AB system (\citealt{oke1983}).

\section{The Q\lowercase{z}5 survey} 
\label{2}

\subsection{Sample}
\label{2.1}

The sample of Qz5 is composed of two different subsets of QSOs at $z\sim5$. The first subset is composed of 22 QSOs from SDSS BOSS (DR9 \& DR10; \citealt{ahn2014,paris2017}) that were observed with Keck/ESI. Because the original goal of the Keck program was to measure the metallicity of DLAs at $z>4.7$, some of these 22 QSOs were originally selected based on the pre-identification of DLAs in the BOSS spectra (\citealt{noterdaeme2012}). However, early in the program it became apparent that DLA candidates identified in the \citet{noterdaeme2012} catalog are false positives due to (1) the increasing density of the \lya\ forest and (2) the low S/N of BOSS spectra for QSOs at these redshifts \citep{rafelski2014}. Hence, the program pivoted to targeting QSOs from BOSS at $4.7<z<5.7$ based solely on redshift and magnitude (i.e.; without any DLA pre-identification). Since none of the non-proximate DLA candidates identified in \citet{noterdaeme2012} are real DLAs or subDLAs \citep{rafelski2014}, we treat all 22 QSOs as an unbiased sample. The properties of these QSOs observed with Keck/ESI are presented in Table \ref{table:sample}.

The second subset of QSOs is based on optical and mid-infrared photometric selection and low spectral resolution follow-up of targets originally identified in the Wide-field Infrared Survey Explorer (WISE; \citealt{WISE}), as described in \citet{wang2016c} and \citet{yang2016}. We observed 41 of these QSOs with VLT/X-SHOOTER (\citealt{x-shooter}) based on their magnitudes and redshifts. The properties of these QSOs are also summarized in Table \ref{table:sample}. In total, the sample of this work is composed of 63 QSOs at $4.7<z<5.7$, and is considered unbiased with regard to the presence of DLAs.

\startlongtable
\begin{deluxetable*}{|c|c|c|c|c|c|c|c|c|}
\label{table:sample}
    \tabletypesize{\normalsize}
    \tablecaption{QSO sample. Columns list the QSO name (1), ra (2), dec (3), QSO z-band magnitude from SDSS (4), QSO redshift measured by Qz5 (5), instrument (6), exposure time (7), median S/N per pixel in the \lya\ forest at the wavelengths of the search path (8), and peak S/N per pixel in the vicinity of the \lya\ emission from the QSO (9).}
    \setlength{\tabcolsep}{0.13in}
    \tablehead{QSO & ra & dec & $m_z$ & $z_{em}$ & Instrument & $t_{exp}$~[s] & S/N$^{1}$ & S/N$^{2}$}
    \startdata
    J0007+0041 & 00:07:49.17 & 00:41:19.62 & 19.84 & 4.76 & ESI & 1800 & 2.2 & 26.1 \\
    J0017-1000 & 00:17:14.67 & -10:00:55.41 & 19.61 & 4.99 &  XSH & 2400 & 2.1 & 37.9 \\
    J0025-0145 & 00:25:26.84 & -01:45:32.50 & 17.86 & 5.06 & XSH & 600 & 2.6 & 48.1 \\
    J0054-0109 & 00:54:21.43 & -01:09:21.67 & 19.57 & 5.03 & ESI & 3600 & 3.5 & 26.0 \\
    J0108+0711 & 01:08:06.59 & 07:11:21.27 & 19.58 & 5.54 & XSH & 2400 & 2.1 & 31.1 \\
    J0115-0253 & 01:15:46.27 & -02:53:12.23 & 19.59 & 5.06 & XSH & 1800 & 1.6 & 27.2 \\
    J0116+0538 & 01:16:14.31 & 05:38:17.59 & 19.27 & 5.33 & XSH & 1800 & 2.3 & 46.0 \\
    J0131-0321 & 01:31:27.35 & -03:21:00.08 & 18.00 & 5.19 & XSH & 600 & 2.0 & 30.6 \\
    J0155+0415 & 01:55:33.28 & 04:15:06.74 & 19.28 & 5.33 & XSH & 2400 & 2.3 & 30.8 \\
    J0216+2304 & 02:16:24.16 & 23:04:09.47 & 19.50 & 5.21 & XSH & 1800 & 1.5 & 26.3 \\
    J0221-0342 & 02:21:12.62 & -03:42:52.31 & 19.56 & 5.01 & XSH & 1800 & 2.4 & 50.8 \\
    J0241+0435 & 02:41:52.92 & 04:35:53.45 & 19.56 & 5.18 & XSH & 1800 & 1.7 & 43.8 \\
    J0251+0333 & 02:51:21.33 & 03:33:17.39 & 19.15 & 4.99 & XSH & 1200 & 3.5 & 53.8 \\
    J0306+1853 & 03:06:42.51 & 18:53:15.82 & 17.66 & 5.33 & XSH & 1200 & 6.6 & 58.2 \\
    J0338+0021 & 03:38:29.32 & 00:21:56.23 & 19.76 & 4.99 & XSH & 2400 & 3.3 & 29.2 \\
    J0747+1153 & 07:47:49.18 & 11:53:52.44 & 18.31 & 5.25 & XSH & 1200 & 2.2 & 64.1 \\
    J0756+4104 & 07:56:18.14 & 41:04:08.59 & 19.90 & 5.09 & ESI & 1800 & 2.4 & 18.9 \\
    J0812+0440 & 08:12:48.82 & 04:40:56.57 & 19.83 & 5.31 & XSH & 4980 & 2.1 & 67.1 \\
    J0835+0537 & 08:35:54.37 & 05:37:53.04 & 19.78 & 5.06 & XSH & 4980 & 2.2 & 30.9 \\
    J0846+0800 & 08:46:27.84 & 08:00:51.72 & 19.64 & 4.99 & XSH & 2400 & 1.7 & 22.5 \\
    J0854+2056 & 08:54:30.37 & 20:56:50.84 & 19.34 & 5.17 & XSH & 1800 & 2.0 & 53.7 \\
    J0902+0851 & 09:02:45.76 & 08:51:15.90 & 19.89 & 5.21 & XSH & 2400 & 1.4 & 51.4 \\
    J0957+0519 & 09:57:27.87 & 05:19:05.26 & 19.64 & 5.18 & ESI & 5400 & 2.4 & 40.4 \\
    J0957+0610 & 09:57:07.68 & 06:10:59.55 & 18.88 & 5.16 & XSH & 1200 & 3.6 & 30.6 \\
    J0957+1016 & 09:57:12.20 & 10:16:18.58 & 19.63 & 5.12 &  XSH & 2400& 0.8 & 20.8 \\
    J1004+2025 & 10:04:44.31 & 20:25:20.03 & 19.69 & 5.00 & XSH & 2400 & 1.6 & 23.2 \\
    J1004+4045 & 10:04:49.59 & 40:45:54.00 & 19.78 & 4.89 & ESI & 3300 & 2.7 & 32.7\\
    J1022+2252 & 10:22:10.04 & 22:52:25.35 & 19.29 & 5.48 & XSH & 2400 & 2.0 & 63.1 \\
    J1028+0746 & 10:28:33.46 & 07:46:18.94 & 19.93 & 5.17 & ESI & 1800 & 2.2 & 27.0 \\
    J1146+4037 & 11:46:57.79 & 40:37:08.59 & 19.30 & 4.98 & ESI & 5400 & 7.2 & 69.2 \\
    J1147-0109 & 11:47:06.42 & -01:09:58.37 & 19.28 & 5.25 & XSH & 1800 & 1.6 & 22.4 \\
    J1200+1817 & 12:00:55.62 & 18:17:32.91 & 19.24 & 4.98 & XSH & 1800 & 4.4 & 41.6 \\
    J1204-0021 & 12:04:41.73 & -00:21:49.54 & 18.97 & 5.08 & ESI & 1200 & 3.4 & 26.7 \\
    J1245+3822 & 12:45:15.46 & 38:22:47.51 & 19.43 & 4.95 & ESI & 1800 & 3.2 & 32.0 \\
    J1332+2208 & 13:32:57.44 & 22:08:35.87 & 19.23 & 5.12 & XSH & 1200 & 3.6 & 52.6 \\
    J1335-0328 & 13:35:56.24 & -03:28:38.29 & 18.97 & 5.68 & XSH & 1200 & 1.3 & 44.0 \\
    J1341+3510 & 13:41:54.02 & 35:10:05.80 & 19.46 & 5.25 & ESI & 2600 & 1.9 & 37.4 \\
    J1345+2329 & 13:45:26.62 & 23:29:49.30 & 18.78 & 5.04 & ESI & 1200 & 1.8 & 30.2 \\
    J1418+3142 & 14:18:39.99 & 31:42:44.07 & 19.23 & 4.91 & ESI & 1800 & 2.7 & 23.5 \\
    J1421+3433 & 14:21:03.83 & 34:33:32.00 & 18.85 & 4.94 & ESI & 2700 & 2.4 & 25.5 \\
    J1423+1303 & 14:23:26.04 & 13:02:57.55 & 19.43 & 5.03 & XSH & 1800 & 3.5 & 38.1 \\
    J1436+2132 & 14:36:05.00 & 21:32:39.26 & 19.28 & 5.22 & XSH & 2400 & 1.8 & 32.4 \\
    J1443+0605 & 14:43:52.94 & 06:05:33.16 & 19.93 & 4.91 & ESI & 1800 & 1.0 & 19.8 \\
    J1443+3623 & 14:43:50.67 & 36:23:15.18 & 19.47 & 5.37 & ESI & 2400 & 1.7 & 35.2 \\
    J1523+3347 & 15:23:45.69 & 33:47:59.41 & 20.16 & 5.35 & ESI & 1800 & 0.9 & 21.4 \\
    J1534+1327 & 15:34:59.76 & 13:27:01.43 & 20.57 & 5.05 & ESI & 1800 & 0.8 & 44.2 \\
    J1536+1437 & 15:36:27.10 & 14:37:17.12 & 19.87 & 4.91 & ESI & 1800 & 1.5 & 26.2 \\
    J1601-1828 & 16:01:11.17 & -18:28:35.07 & 19.54 & 5.03 & XSH & 1800 & 1.4 & 25.4\\
    J1614+2059 & 16:14:47.04 & 20:59:02.85 & 19.74 & 5.06 & ESI & 1200 & 2.5 & 19.0 \\
    J2201+0302 & 22:01:06.63 & 03:02:07.70 & 18.97 & 5.06 & XSH & 1200 & 3.0 & 29.4 \\
    J2202+1509 & 22:02:26.77 & 15:09:52.37 & 18.55 & 5.07 & XSH & 960 & 2.4 & 35.3 \\
    J2207-0416 & 22:07:10.13 & -04:16:56.22 & 19.10 & 5.51 & XSH & 1200 & 1.1 & 44.7 \\
    J2216+0013 & 22:16:44.02 & 00:13:48.14 & 19.93 & 5.01 & ESI & 1200 & 1.8 & 19.3 \\
    J2220-0101 & 22:20:18.50 & -01:01:47.08 & 20.13 & 5.62 & ESI & 2400 & 0.7 & 24.1 \\
    J2225+0330 & 22:25:14.38 & 03:30:12.52 & 19.64 & 5.25 & XSH & 2400 & 0.9 & 21.7 \\
    J2226-0618 & 22:26:12.42 & -06:18:07.35 & 18.77 & 5.09 & XSH & 600 & 1.5 & 24.5 \\
    J2228-0757 & 22:28:45.14 & -07:57:55.30 & 19.38 & 5.15 & XSH & 3600 & 1.5 & 41.7 \\
    J2312+0100 & 23:12:16.45 & 01:00:51.58 & 20.88 & 5.08 & ESI & 2400 & 0.7 & 29.9 \\
    J2325-0553 & 23:25:36.64 & -05:53:28.42 & 19.19 & 5.20 & XSH & 1200 & 1.4 & 21.7 \\
    J2330+0957 & 23:30:08.71 & 09:57:43.73 & 19.75 & 5.25 & XSH & 2400 & 1.3 & 17.0 \\
    J2344+1653 & 23:44:33.50 & 16:53:16.58 & 18.60 & 5.00 & XSH & 600 & 1.9 & 25.5 \\
    J2351-0459 & 23:51:24.31 & -04:59:07.30 & 19.59 & 5.26 & XSH & 2400 & 1.9 & 30.7\\
    J2358+0634 & 23:58:24.05 & 06:34:37.48 & 19.52 & 5.27 & XSH & 2400 & 2.7 & 30.3 \\
    \enddata
\end{deluxetable*}

\subsection{Observations and data}
\label{2.2}

The Keck/ESI data were obtained in 2013 January, 2013 May, and 2013 August with the $\sim 0.\arcsec 75$ slit. The seeing varied between 0.6-1.1\arcsec. This configuration corresponds to a spectral resolution of $R\sim7000$, which is equivalent to $\sim44$~km~s$^{-1}$ (FWHM). The Keck/ESI data data were dithered along the slit with a three-point dither pattern to decrease the effects of fringing, although not all observations included all three dither positions.

The VLT/X-SHOOTER data were obtained in service mode through the programs 098.A-0111 and 0100.A-0243 (PI: M.~Rafelski). The observations were conducted between October 2017 and March 2018 with the $0\farcs9$ slit in the VIS arm and $0\farcs6$ slit in the NIR arm. For the VIS ($\lambda \sim 0.5 - 1~\mu \rm m$) arm, this corresponds to $R\sim8900$ and $\sim34$~km~s$^{-1}$ (FWHM). For the NIR arm ($\lambda \sim 1-2.5~\mu \rm m$), the data feature $R\sim8100$, which is equivalent to $\sim37$ km s$^{-1}$ (FWHM). We used 1x2 binning for the VIS arm data, which still sufficiently samples the PSF with 3 pixels per resolution element. 

\subsection{Data reduction}
\label{2.3}

To process the Keck/ESI data, we turned to the software \texttt{XIDL} developed by J.X.~Prochaska. We utilized the package \texttt{ESIRedux} with optimal extraction (\citealt{prochaska2003}) that follows the same methodology described in \citet{rafelski2012}.

To process the VLT/X-SHOOTER data, we used a set of custom built IDL software codes written by G.~Becker. Previous testing of these codes on quasar spectra from the XQ-100 survey \citep{lopez16} revealed improved sky subtraction over the standard ESO pipeline \citep{modigliani10}, especially for the near-infrared spectra. One of the main differences between these two pipelines is the nodding strategy. The IDL routine used in this work does not implement nodding for the subtraction of the sky background, avoiding a factor of \(\sqrt{2}\) increase in the noise during reduction. 

Strong Telluric absorption lines affect both the optical and near-infrared spectral ranges. After each of the quasar observations, VLT/X-SHOOTER also carried out associated observations of a hot B-type star to allow us to model the strengths of the atmospheric absorption lines. To correct for the pressure, temperature and airmass-dependent strengths, we used the ESO \texttt{Molecfit} tool \citep{smette15} to model the transmission in the Telluric standard star spectra and correct the extracted quasar spectra for the atmospheric absorption lines.

The final 1-dimensional spectra were extracted, combined, and flux calibrated using observations of spectro-photometric standard stars obtained during the night of the science observations. Wavelength calibration corrected all spectra into vacuum, Heliocentric frame. We found a systematic wavelength shift in the wavelength calibrated VLT/X-SHOOTER spectra\footnote{\url{https://www.eso.org/sci/facilities/paranal/instruments/xshooter/doc/XS_wlc_shift_150615.pdf}}. We determined the magnitude of this wavelength shift through a cross-correlation between the reduced spectra and the Cerro Paranal Advanced Sky Model\footnote{\url{https://www.eso.org/observing/etc/bin/gen/form?INS.MODE=swspectr+INS.NAME=SKYCALC}}. All shifts were found to be within $\lesssim 1$~\AA\ and were used to correct the wavelength solution of the reduced data products. This process will be outlined in greater detail in a companion paper that focuses on the metal line absorption (Wisz et al.~in prep). Appendix \ref{appendix1} provides the hyperlink to download the data and shows the spectra for the 63 QSOs.

\begin{figure}[b]
\centering
\includegraphics[width=3.35in]{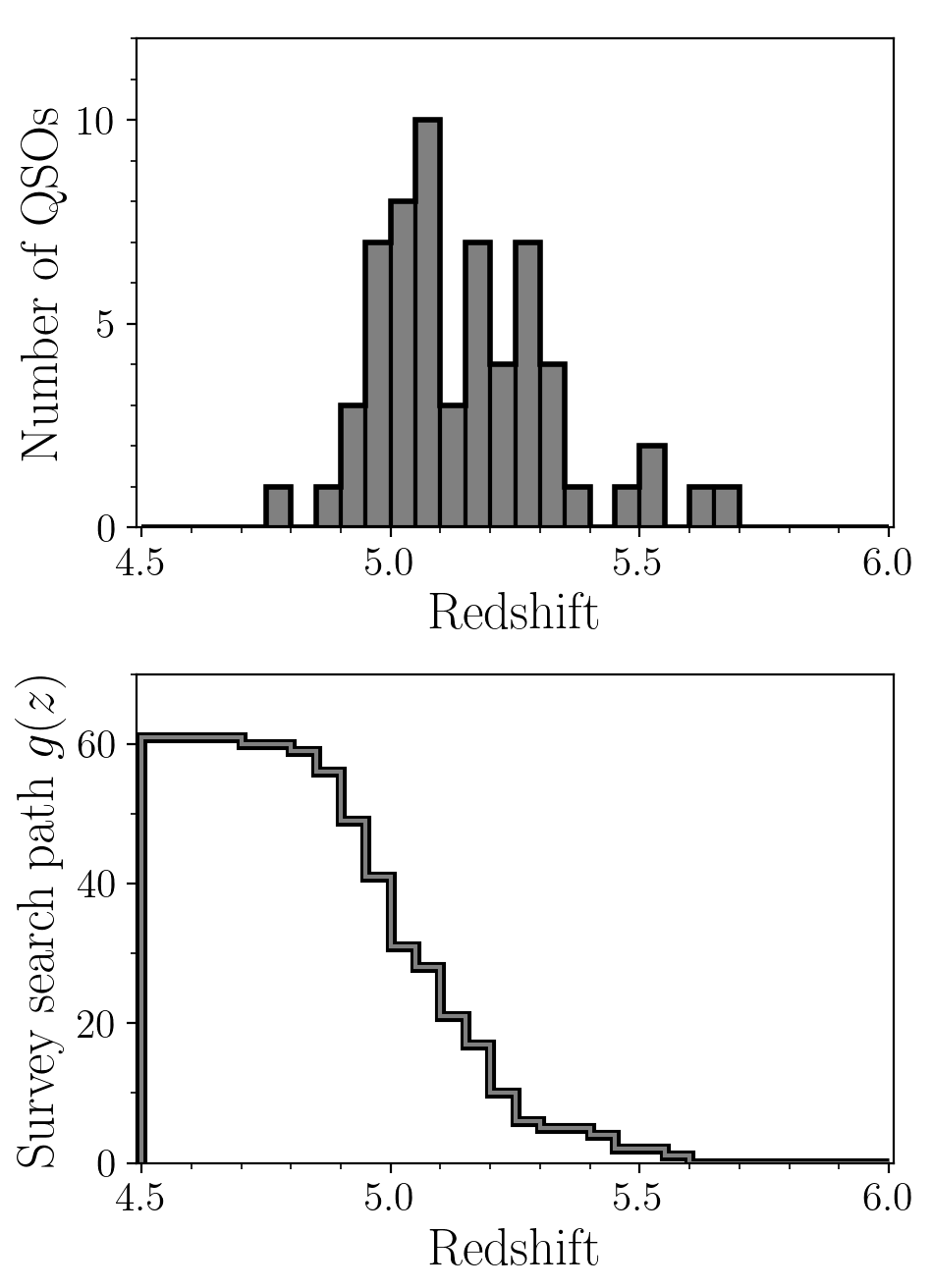}
    \caption{The QSO sample comprising Qz5. \textbf{Top:} Histogram of the redshifts of the 63 QSOs in our sample. We carried out high resolution spectroscopic observations of these QSOs in order to identify DLAs and constrain \rhoHI\ at $z\sim 5$. \textbf{Bottom:} Number of sightlines available for the identification of DLAs (i.e., $g(z)$, the survey search path of Qz5) as a function of redshift. The search path remains high out to $z\sim 4.9$ before dropping toward $z\sim 5.5$.}
   \label{fig:sample}
\end{figure}

\subsection{Emission redshifts and survey search path}
\label{2.4}
We utilized the reduced spectra to remeasure the emission redshifts of the QSOs. To this end, we designed an algorithm that maximizes the cross-correlation between the observed spectra and an estimate of the QSO continuum obtained with \texttt{Quasar Factor Analysis} (\citealt{sun2023}). The QSO redshifts are presented in Table \ref{table:sample} and the redshift distribution of the sample is shown in Figure \ref{fig:sample}. The typical uncertainties on the redshifts are $z_{err} = 0.05$, and we found our redshifts and those measured by SDSS and/or \citet{wang2016c} to be within $\Delta z = 0.08$ in all cases. 

In Figure \ref{fig:sample}, we also show the survey search path or $g(z)$ of Qz5. Because all 63 quasars in Qz5 have emission redshifts $z_{em}>4.7$, the available path length between $z\sim 4.5-4.7$ is 63. This path length decreases toward lower and higher redshift because fewer QSOs are available to search for Ly$\alpha$ absorbers. Therefore, and because the goal of our study is to constrain \rhoHI\ at $z\sim 5$, we limit our analysis to the redshift range $z= 4.5 - 5.6$. It is apparent in Figure \ref{fig:sample} that $g(z)$ of Qz5 is at its highest out to $z\sim 4.9$, with a rapid decrease as we approach $z\sim 5.5$.

\section{Methodology} 
\label{3}

\subsection{Analysis}
\label{3.2}

Because of the high spectral resolution, adequate signal-to-noise, and the moderate sample size of our dataset, we were able to search for DLAs and subDLAs through visual inspection without the need for an automated algorithm. Effectively, we selected all regions with widths greater than 10~\AA\ in the \lya\ forest of every QSO that are consistent with zero flux. Our search yielded a sample of 39 high-\NHI\ absorber candidates in the 63 QSOs.

\begin{figure*}
\centering
\includegraphics[width=7.1in]{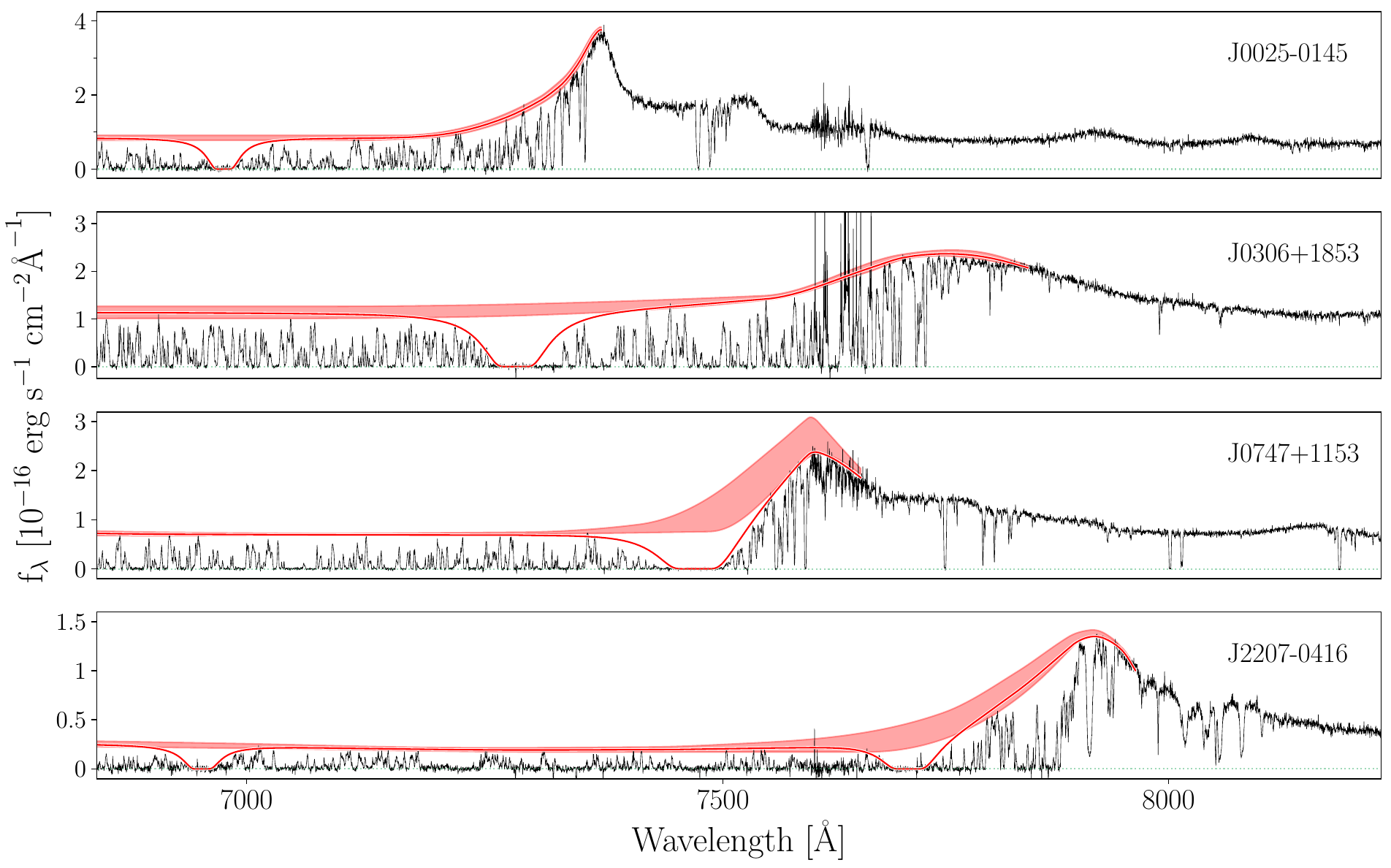}
    \caption{Spectra of the 4 QSOs featuring the 5 DLAs detected in Qz5. The QSO identifiers are shown in the top right corner of every panel. All these QSOs were observed with VLT/X-SHOOTER. Also plotted are the estimated QSO continuum fluxes blueward of \lya\ (red shading) and the associated Voigt profile fits to the DLAs used to determine their \NHI\ (red lines). Details on how these continua were determined are presented in Section \ref{3.2}.}
   \label{fig:spectra}
\end{figure*}

Distinguishing between DLAs and subDLAs (i.e., measuring \NHI\ for all candidates) requires estimation of the QSO continuum blueward of the Ly$\alpha$ emission peak. We note that the \lya\ forest is very dense at these redshifts, and thus the standard practice of only fitting \NHI\ to the peaks of the \lya\ forest could slightly overestimate \NHI. Therefore, we fit two continua in the \lya\ forest per target. The first fits were set to closely align with the peaks of the \lya\ forest (as typically done; e.g. \citealt{rafelski2012}). To account for possible \lya\ absorption, the second fits were set to higher flux density values (typically by 10-20\%). These fits aim to encompass the maximum QSO continuum level allowed by template QSO spectra at lower redshifts, including the \lya\ emission peak (see Figure \ref{fig:spectra} for examples). The two continuum fits were averaged to determine the best continuum estimate, which was then used to normalize the QSO spectrum for the characterization of \HI\ absorbers.

\begin{deluxetable*}{cccccccc}
\tabletypesize{\normalsize}
\label{table:results}
\tablecaption{DLAs identified in Qz5. The columns correspond to QSO identifier (1), ra (2), dec (3), emission redshift (4), absorption redshift (5), notes (6), and log\NHI (7).}
\setlength{\tabcolsep}{0.17in}
\tablehead{
\colhead{QSO} & \colhead{ra} & \colhead{dec} & \colhead{z$_{em}$} & \colhead{z$_{abs}$} & \colhead{Notes} & \colhead{log\NHI~[cm$^{-2}$]} } 
\startdata 
J0007+0041 & 00:07:49.17 & 00:41:19.62 & 4.76 & 4.7330 & \xmark: Proximate DLA & 20.6 $\pm$ 0.3\\
J0025-0145 & 00:25:26.84 & -01:45:32.50 & 5.06 & 4.7389 & \cmark: DLA & 20.3 $\pm$ 0.15\\
J0306+1853 & 03:06:42.51 & 18:53:15.82 & 5.33 & 4.9866 & \cmark: DLA & 20.9 $\pm$ 0.15 \\
J0747+1153 & 07:47:49.18 & 11:53:52.44 & 5.25 & 5.1447 & \cmark: DLA & 21.1 $\pm$ 0.25\\
J1345+2329 & 13:45:26.62 & 23:29:49.30 & 5.04 & 5.0060 & \xmark: Proximate DLA & 21.1 $\pm$ 0.1 \\
J1436+2132 & 14:36:05.00 & 21:32:39.26 & 5.22 & 5.1780 & \xmark: Proximate DLA & 20.7 $\pm$ 0.25\\
J2207-0416 & 22:07:10.13 & -04:16:56.22 & 5.51 & 4.7220 & \cmark: DLA & 20.4 $\pm$ 0.2 \\
... & ... & ... & ... &  5.3374 & \cmark: DLA & 20.8 $\pm$ 0.15 \\
\enddata
\end{deluxetable*}

This best continuum estimate was utilized to visually determine the best fit \NHI\ (to which we will refer as simply \NHI) for all candidate absorbers. The uncertainties (to which we will refer as $\Delta$\NHI) were determined by visually fitting the DLA at both the lower and higher continuum levels to fully encompass the uncertainty on \NHI. We note that $\Delta$\NHI\ is dominated by the uncertainties on the QSO continuum, and thus our uncertainties are conservative. Of the 39 high-\NHI\ absorbers, 8 were found to satisfy the DLA column density threshold (i.e., \NHI\,$\geqslant 10^{20.3}$~cm$^{-2}$) and 31 were best classified as subDLAs (i.e., $10^{19}\leqslant$~\NHI~[cm$^{-2}$]~$<10^{20.3}$). These 8 DLAs were fitted by three authors independently (G.~A.~Oyarz\'un, M. Rafelski, and F.~Ozyurt), and the recovered values for \NHI\ were within $\Delta$\NHI\ in every case.

\begin{figure*}
\centering
\includegraphics[width=7.1in]{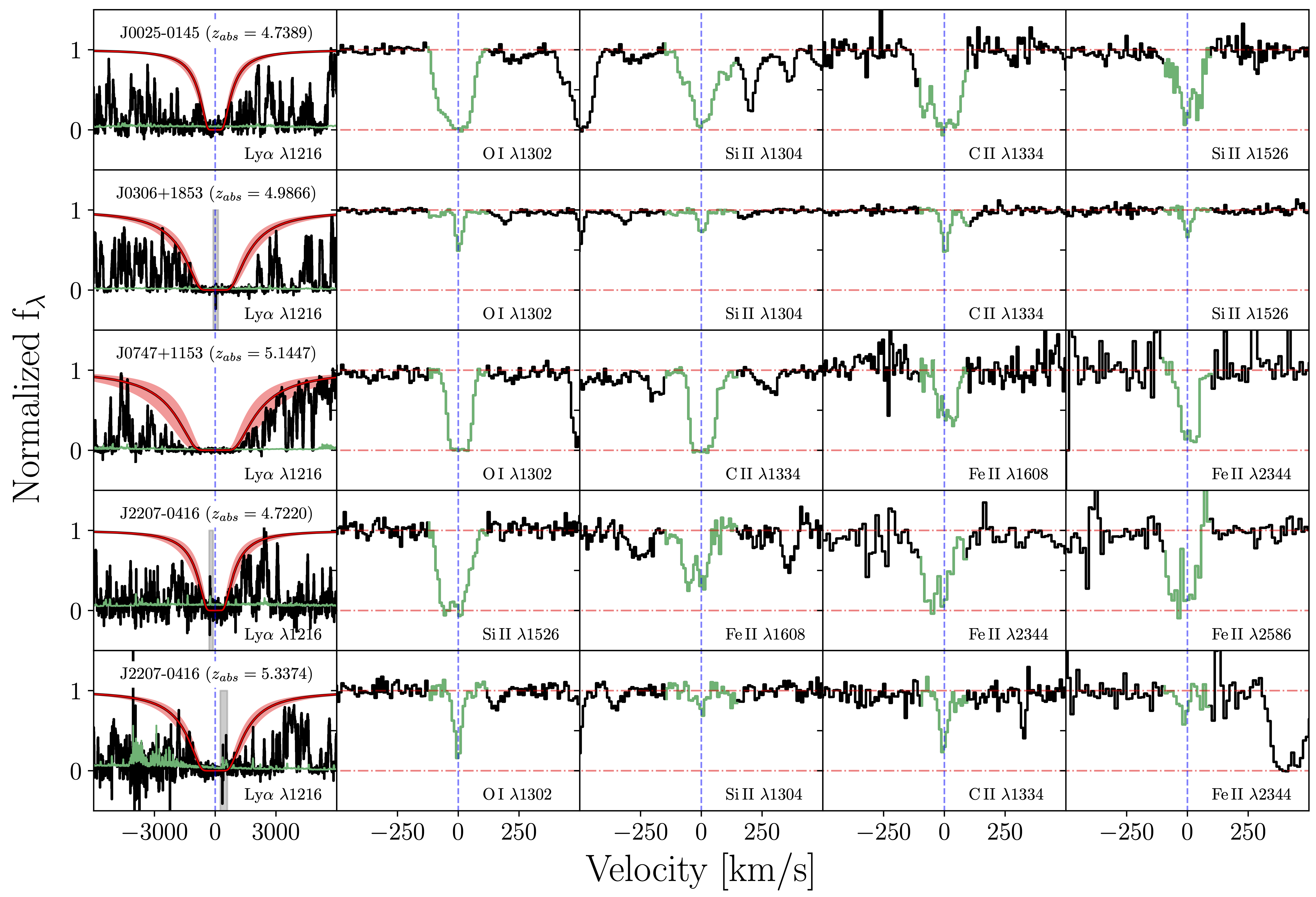}
    \caption{Detailed spectroscopic view of the 5 DLAs at $z\sim 5$ identified in Qz5. The rows show the 5 different DLAs. The leftmost column shows the QSO spectrum zoomed-in on the \HI\ absorption of the DLA (black), the error on the flux (green), and the Voigt fit used to estimate \NHI\ (red). Residuals from sky subtraction are shaded in gray. The 4 rightmost columns show the continuum normalized QSO spectrum (black) zoomed-in on 4 different metal lines (green) that were used to determine the redshift of the \HI\ gas absorber.}
   \label{fig:metals}
\end{figure*}

The DLA sample was subject to further processing and analysis, including continuum fitting redward of \lya\ emission from the QSO. The best fit continuum, which was determined with the \texttt{lt\_continuumfit} routine of the \texttt{linetools} package\footnote{\url{https://zenodo.org/records/168270}} (\citealt{linetools}) was used to normalize the QSO spectrum for the characterization of metal line absorption and subsequent absorption redshift determination. A wide range of metal lines was used in this process, in most cases including $\mathrm{O\,I}~\lambda 1302$, $\mathrm{Si\,II}~\lambda 1304$, $\mathrm{C\,II}~\lambda 1334$, $\mathrm{Si\,II}~\lambda 1526$, and $\mathrm{Fe\,II}~\lambda 1608$. We then used the routine \texttt{pyigm\_fitdla} of the \texttt{linetools} package to derive redshifts from the metal lines for the 8 DLAs. No DLA candidates without associated metal lines were identified. The properties of the 8 DLAs found are summarized in Table \ref{table:results}.

The redshifts were used to exclude all proximate absorbers --- i.e., all absorbers within $<5000$~km~s$^{-1}$ of \lya\ emission from the QSO --- because they can no longer be considered an intervening gas cloud that is not associated with the QSO (\citealt{prochaska2008,ellison2010,ellison2011}). Three of the 8 DLAs and two of the 31 subDLAs were found to be proximate. After their removal from the sample, we obtained a total of 5 intervening DLAs and 29 intervening subDLAs. The \NHI\ fits for the DLAs are shown in Figure \ref{fig:spectra}, and some of the metal lines used to determine their redshifts are plotted in Figure \ref{fig:metals}.  All absorbers (i.e. both DLAs and subDLAs) that were identified in this analysis are presented in Appendix \ref{appendix2}.

\subsection{Formalism for measuring \rhoHI}
\label{3.3}

We can use our dataset of detected and characterized DLAs to constrain the \HI\ column density distribution. This distribution function is typically denoted as $f(\mbox{\NHI}, X)$, where \NHI\ is the \HI\ column density and $X$ is the absorption distance (\citealt{bahcall-peebles1969,lanzetta1991,wolfe2005}). The latter is related to the geometry of the Universe by the relation
\begin{equation}
dX = \frac{H_0}{H(z)}(1+z)^2 dz.
\end{equation}

In practice, we measured $f(\mbox{\NHI}, X)$ in bins of \NHI\ and $X$, i.e.,
\begin{equation}
f(\mbox{\NHI}, X) = \dfrac{\rm N_{\rm DLA} (\mbox{\NHI}, \mbox{\NHI} \pm \delta\mbox{\NHI}/2) }{\delta X}, 
\end{equation}
where $\delta$\NHI\ and $\delta X$ represent the width of the bins in \NHI\ and $X$, respectively. $\rm N_{\rm DLA}$ denotes the number of DLAs at the given \NHI\ and $X$ bin (\citealt{prochaska2005}).

Measurements of $f(\mbox{\NHI}, X)$ can be utilized to infer $\ell_{\rm DLA}(X)$, the DLA incidence rate. In equation form,   
\begin{equation}
\label{l(x)}
\ell_{\rm DLA}(X)\, dX = \dfrac{d\mbox{\NHI}}{dX}\, dX = \int^{\infty}_{N_{min}} f(\mbox{\NHI}, X)\, d\mbox{\NHI} \,dX,
\end{equation}
where $N_{min}$ corresponds to the lower limit for the column density of DLAs, i.e., $N_{min} = 10^{20.3}$~cm$^{-2}$. By definition, the DLA incidence rate captures the total number of high \NHI\ absorbers in the Universe at a given redshift. Because of this, $\ell_{\rm DLA}(X)$ is independent of systems with \NHI~$<10^{20.3}$~cm$^{-2}$ (i.e.; subDLAs) that are more difficult to quantify in observations with high completeness.

The \HI\ mass density associated with DLAs is captured in the first moment of $f(\mbox{\NHI}, X)$, i.e., 
\begin{equation}
\label{eq:rho_dla}
\rho_{\mbox{\tiny DLA}}(X)\,dX = \dfrac{H_0 m_{\mbox{\scriptsize H}}}{c} \int^{\infty}_{N_{min}} \mbox{\NHI} f(\mbox{\NHI}, X)\, d\mbox{\NHI}\,dX. 
\end{equation}
Obtaining Equation (\ref{eq:rho_dla}) required writing the mass density of \HI\ in DLAs as 

\begin{equation}
\rho_{\mbox{\tiny DLA}} = n_{\mbox{\tiny DLA}} m_{\mbox{\tiny DLA}}
\end{equation}
and 
\begin{equation}
\rho_{\mbox{\tiny DLA}}(X)\,dX = n_{\mbox{\tiny DLA}}(X) A(X) m_{\mbox{\scriptsize H}} \mbox{\NHI}\,dX, 
\end{equation}
where $n_{\mbox{\tiny DLA}}$ is the comoving number density of DLAs, $m_{\mbox{\tiny DLA}}$ is the \HI\ mass per DLA, $A(X)$ is the proper absorption cross section, and $m_{\mbox{\scriptsize H}}$ is the mass of the hydrogen atom.  

Upon inclusion of the critical density of the Universe ($\rho_{\mbox{\scriptsize \sl \rm crit, 0}}$), we obtain
\begin{equation}
\begin{split}
&\dfrac{\rho_{\mbox{\tiny DLA}}(X)}{{\rho_{\mbox{\scriptsize \sl \rm crit, 0}}}}\,dX = \\*
\dfrac{H_0}{c} \dfrac{m_{\mbox{\scriptsize H}}}{\rho_{\mbox{\scriptsize \sl \rm crit, 0}}} \int^{\infty}_{N_{min}} \mbox{\NHI} & f(\mbox{\NHI}, X)\, d\mbox{\NHI}\,dX =\\*
& \Omega_{\mbox{\tiny DLA}}(X)\,dX.
\end{split}
\end{equation}

Converting between $\rho_{\mbox{\tiny DLA}}$ and \rhoHI\ requires knowledge of the shape of $f(\mbox{\NHI}, X)$ toward lower \NHI. For instance, observations of $f(\mbox{\NHI}, X)$ at $z\sim 2-4$ (e.g. \citealt{omeara2007,noterdaeme2009,prochaska2010,zafar2013}) have found that the contribution from subDLAs to \rhoHI\ is $10-20\%$. For instance, \citet{crighton2015} adopted a contribution from subDLAs of $\approx 17\%$ at $z\sim 5$, corresponding to a correction factor
\begin{equation}
\mbox{\rhoHI} = \delta_{\scriptsize \HI} \, \rho_{\mbox{\tiny DLA}} = 1.2 \, \rho_{\mbox{\tiny DLA}},
\end{equation}

With studies showing no evidence of an evolution in $\delta_{\scriptsize \HI}$ across the $z\sim 2-4$ range (e.g. \citealt{berg2019}), the choice of $\delta_{\scriptsize \HI} = 1.2$ is justified only up to $z\sim 4$. In this paper, we quantify $\delta_{\scriptsize \HI}$ directly from the subDLA sample (Section \ref{4}).

We can then write our final equation that quantifies \rhoHI\ from our observations as
\begin{equation}
\begin{split}
&\dfrac{\mbox{\rhoHI}(X)}{{\rho_{\mbox{\scriptsize \sl \rm crit, 0}}}}\,dX = \\*
\dfrac{8\pi G}{3H_0} \dfrac{m_{\mbox{\scriptsize H}}}{c} \delta_{\scriptsize \HI} \int^{\infty}_{N_{min}} \mbox{\NHI} & f(\mbox{\NHI}, X)\, d\mbox{\NHI}\,dX = \\*
& \mbox{\OmHI}(X)\,dX,
\end{split}
\end{equation}
which was computed by discretely sampling the integrand in narrow \NHI\ bins ($d\mbox{\NHI} = 10^{20}$~cm$^{-2}$) and integrating over redshift intervals. 

\begin{figure*}
\centering
\includegraphics[width=7.1in]{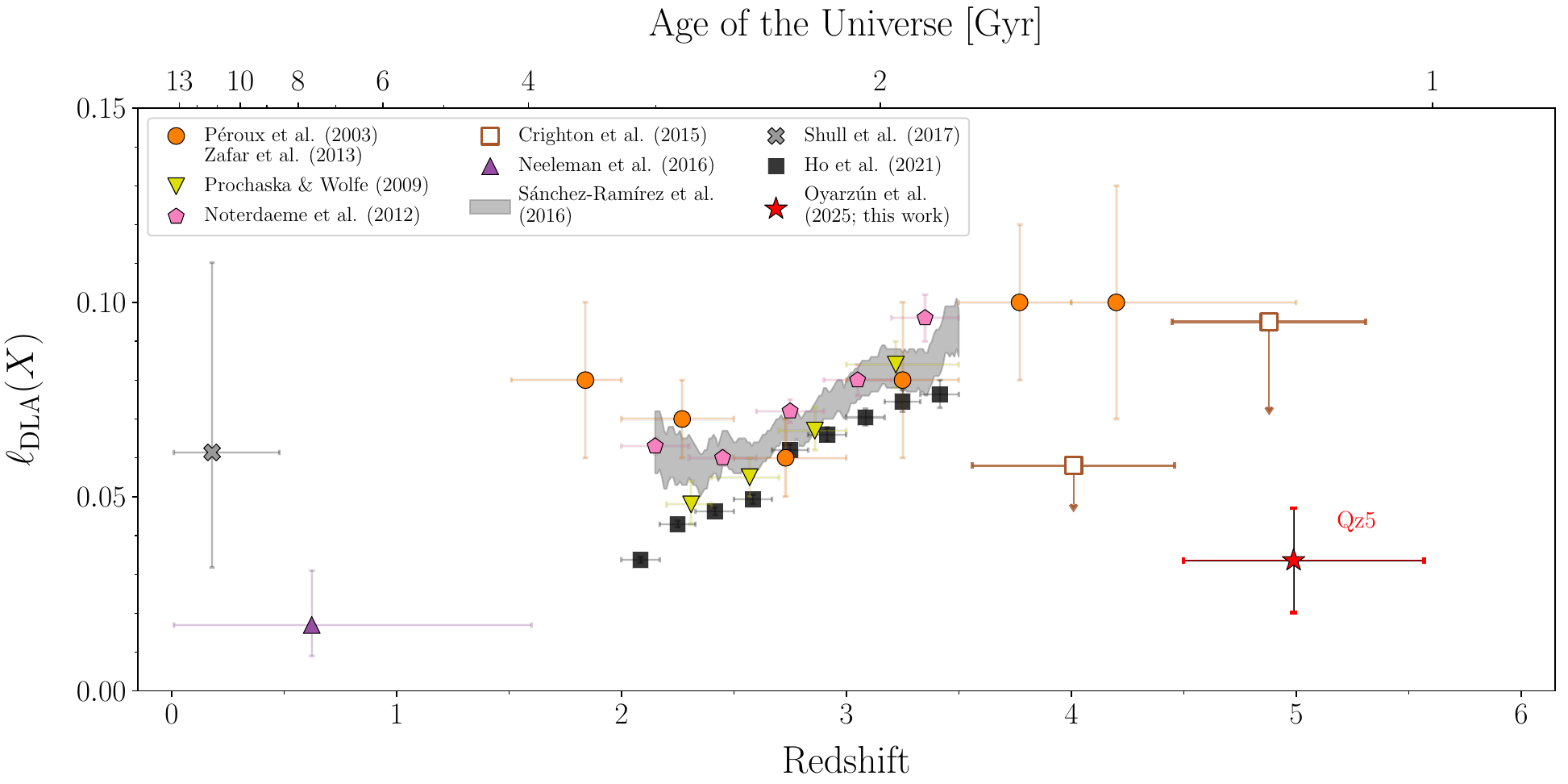}
    \caption{The DLA incidence rate at $z\sim 5$ derived with Qz5, along with estimates from other DLA surveys from the literature. Plotted are the measurements by \citet{peroux2003} and \citet[orange circles]{zafar2013}, \citet[yellow triangles]{prochaska-wolfe2009}, \citet[pink pentagons]{noterdaeme2012}, \citet[brown squares]{crighton2015}, \citet[purple triangle]{neeleman2016}, \citet[gray shading]{sanchez-ramirez2016}, \citet[gray crosses]{shull2017} and \citet[black squares]{ho-bird-garnett2020}. The measurements by \citet{crighton2015} are shown as open square upper limits due to systematic biases (Section \ref{5.1}). Our data point at $z\sim 5$ (red star) is indicative of a much lower DLA incidence rate at $z\sim 5$ than at lower redshifts. We follow the convention that the error bar on the x-axis represents the width of the redshift bin encompassing the sample.}
   \label{fig:l_x}
\end{figure*}

\begin{figure*}
\centering
\includegraphics[width=7.1in]{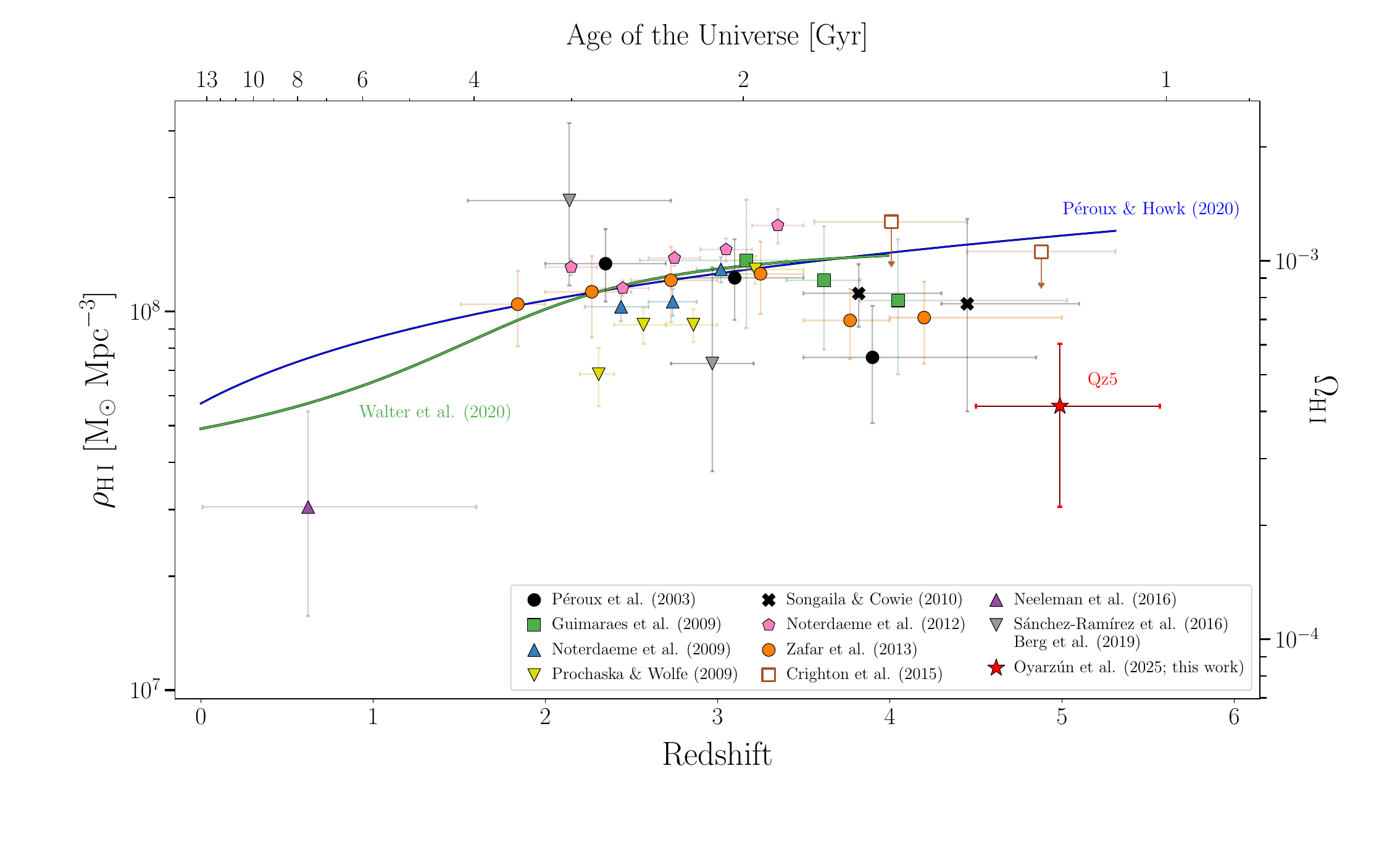}
    \caption{The cosmic \HI\ mass density as a function of redshift as constrained by different DLA surveys. Included in this figure are the measurements by \citet[black circles]{peroux2003}, \citet[green squares]{guimaraes2009}, \citet[blue triangles]{noterdaeme2009}, \citet[yellow triangles]{prochaska-wolfe2009}, \citet[black crosses]{songaila-cowie2010}, \citet[pink pentagons]{noterdaeme2012}, \citet[orange circles]{zafar2013}, \citet[brown squares]{crighton2015}, \citet[purple triangle]{neeleman2016}, and \citet[gray triangles]{sanchez-ramirez2016}. Measurements at high redshift obtained with low spectral resolution data are shown as open symbol upper limits due to systematic biases (\citealt{crighton2015}; Section \ref{5.1}). The blue line shows the fit by \citet{peroux-howk2020} and the green line represents the fit by \citet{walter2020}. The Qz5 data point at $z\sim 5$ is plotted as a red star. The error bar on the x-axis represents the width of the redshift bin. All constraints and fits in this figure adopt a contribution from lower \NHI\ systems of $\delta_{\scriptsize \HI} = 1.2$, as found at $z\sim 2-4$. The only exception is Qz5, where we measure $\delta_{\scriptsize \HI} = 1.4$ (increasing our value for \rhoHI\ in comparison).}
   \label{fig:rho_dla}
\end{figure*}

\begin{figure*}
\centering
\includegraphics[width=7.1in]{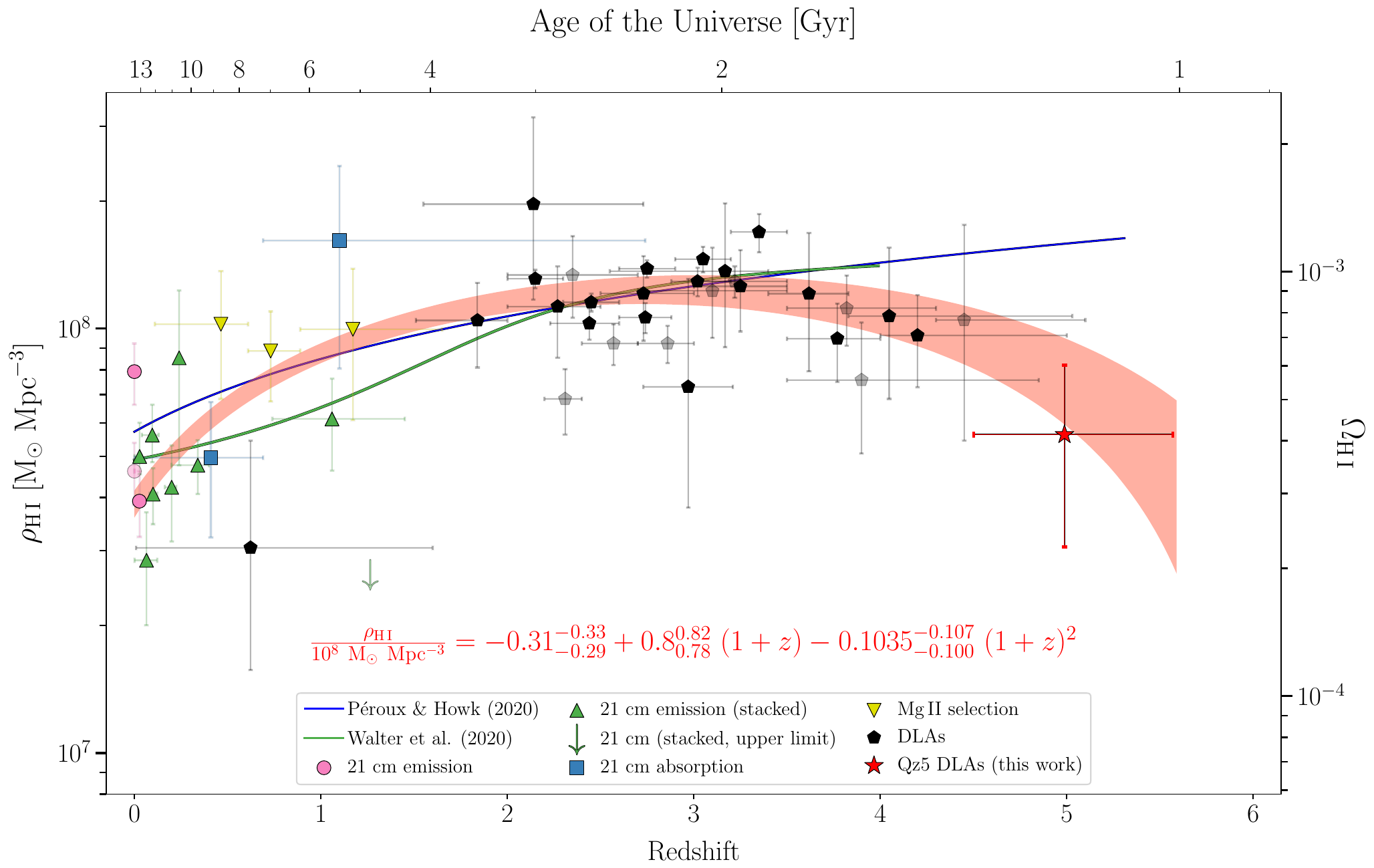}
    \caption{The cosmic \HI\ mass density as a function of redshift according to different observational probes. Different symbols correspond to different techniques for constraining \rhoHI. Constraints from 21~cm emission are plotted as pink circles and include the measurements by \citet{zwaan2005,braun2012}, and \citet{jones2018}. Results from 21~cm emission stacking are plotted as green upward-facing triangles and were originally obtained by \citet{lah2007,delhaize2013,rhee2013,hoppman2015,bera2019} and \citet{chowdhury2020}. The 21~cm emission stacking upper limit (downward pointing black arrow) was measured by \citet{kanekar2016}. The measurements of \rhoHI\ from 21~cm absorption plotted as blue squares correspond to the work by \citet{grasha2020}. The yellow downward-facing triangles show the constraints from \citet{rao2017} using Mg\,II absorbers. The black pentagons show estimates from DLA statistics (see Figure \ref{fig:rho_dla}) after excluding data points affected by resolution systematics (\citealt{noterdaeme2009,prochaska-wolfe2009,crighton2015,sanchez-ramirez2016}). Our data point at $z\sim 5$ is plotted as a red star. Bayesian analysis reveals that the data in this figure is better reproduced by models with a peak in \rhoHI\ (red shading) instead of models where \rhoHI\ monotonically increases with redshift (e.g. \citealt{peroux-howk2020} or \citealt{walter2020}). Due to statistical interdependence between some of the data, only the fully opaque datapoints were utilized for constraining the model posterior.}
   \label{fig:rho_all}
\end{figure*}

We should also note that there is a distinction to be made between \rhoHI\ and the cosmic mass density of gas (often referred to as $\rho_{\rm gas}$). Work often determine $\rho_{\rm gas}$ by implementing a correction for the abundance of Helium, i.e., 
\begin{equation}
\rho_{\rm gas} = \delta_{\scriptsize \rm He}\, \mbox{\rhoHI}.
\end{equation}
Typically adopted is a correction factor of $\delta_{\scriptsize He} \approx 1.3$ (e.g. \citealt{peroux-howk2020,walter2020}).

\section{Results} 
\label{4}

In Qz5, we detect 5 non-proximate DLAs at $z\sim 5$ (Table \ref{table:results}), which corresponds to $<0.1$ DLAs per sightline. This fraction is considerably lower than the $\sim 0.3$ recovered by other surveys at $z\gtrsim 3.5$ (e.g. \citealt{crighton2015}). This is quantitatively apparent in the DLA incidence rate, which has a value of $\ell_{\rm DLA}(X) = 0.034_{0.02}^{0.05}$ for Qz5 (see Table \ref{table:simulated_crighton}) and is plotted alongside estimates from other DLA studies in Figure \ref{fig:l_x}. Here, the uncertainties on $\ell_{\rm DLA}(X)$ were estimated through Monte Carlo simulations where the number of DLAs detected is given by a Poisson distribution. We note that our $\ell_{\rm DLA}(X)$ is inconsistent at the $\sim 4.5\sigma$ level with the mean values from other surveys at $z\sim 4-5$.

Previous work have established that the identification of DLAs in low-resolution and low S/N spectra at $z\gtrsim 3.5$ is unreliable (\citealt{noterdaeme2012,rafelski2012}). For this reason, measurements in Figure \ref{fig:l_x} that are based on low resolution, low S/N data beyond $z\gtrsim 3.5$ are excluded (\citealt{noterdaeme2009,prochaska-wolfe2009,ho-bird-garnett2020,sanchez-ramirez2016}). As we will show in Section \ref{5.1}, these concerns also apply to low resolution, high S/N data. Thus, the measurements by \citet{crighton2015} at $z\sim 4-5$ are plotted as open square upper limits in Figure \ref{fig:l_x}.

Considering only the most reliable values from the literature, Figure \ref{fig:l_x} suggests an evolution in the DLA incidence rate starting with $\ell_{\rm DLA}(X)\sim0.03$ at $z\sim 5$ (this work), peaking around $\ell_{\rm DLA}(X)\sim0.08$ at $z\sim3-4$ (\citealt{peroux2003,prochaska-wolfe2009,noterdaeme2012,zafar2013,sanchez-ramirez2016,ho-bird-garnett2020}), and then decreasing toward $z\sim 0-1$ (\citealt{neeleman2016,shull2017}).

Using the formalism described in Section \ref{3.3}, we also estimated \rhoHI\ at $z\sim 5$ as the addition between the \HI\ mass densities of DLAs and subDLAs.  We obtain \rhoHI~$=0.56^{0.82}_{0.31}\times 10^8~\mbox{M}_{\odot}$~Mpc$^{-3}$ at $z\sim 5$ from Qz5, where the values from each component are reported in Table \ref{table:simulated_crighton}. For the uncertainties, we utilized Monte Carlo simulations that (1) bootstrap over the DLAs and subDLAs detected by Qz5 according to a Poisson distribution and (2) assign the \HI\ column densities from normal distributions with means \NHI\ and standard deviations $\Delta$\NHI.

Our measurement is shown in Figure \ref{fig:rho_dla} along with other estimates from previous DLA campaigns. Because the value of \rhoHI\ is dependent on the value of $H_0$, included in Figure \ref{fig:rho_dla} are only studies that report their choice of cosmological parameters (\citealt{peroux2003,guimaraes2009,prochaska-wolfe2009,noterdaeme2009,noterdaeme2012,songaila-cowie2010,zafar2013,crighton2015,neeleman2016,sanchez-ramirez2016,berg2019}). As we did for $\ell_{\rm DLA}(X)$, we excluded estimates of \rhoHI\ beyond $z\gtrsim 3.5$ that are based on low resolution, low S/N data (i.e.; \citealt{noterdaeme2009,prochaska-wolfe2009,sanchez-ramirez2016}). As in Figure \ref{fig:l_x}, we also show estimates of \rhoHI\ obtained with low resolution, high S/N data at $z\gtrsim 3.5$ as open square upper limits (\citealt{crighton2015}). Also plotted in this figure are the fits to the data by \citet{peroux-howk2020} and \citet{walter2020} after de-correcting for the contribution from Helium ($\delta_{\rm He} = 1.3$).

All estimates that do not measure the subDLA contribution directly (including the fits by \citealt{peroux-howk2020} and \citealt{walter2020}) were plotted assuming a correction factor of $\delta_{\scriptsize \HI} = 1.2$. This factor corresponds to a contribution to \rhoHI\ from low \NHI\ systems of $\approx 17\%$ ($z\sim 2-4$; e.g. \citealt{zafar2013,berg2019}), which is considerably lower than the $\approx 30\%$ that we measure with Qz5 (equivalent to $\delta_{\scriptsize \HI} = 1.44$). However, we note that our estimate of the \HI\ mass density of subDLAs is not as precise as other measurements at $z\sim 2-4$ (\citealt{zafar2013,berg2019}) due to our smaller search path.

Despite the larger $\delta_{\scriptsize \HI}$ at $z\sim 5$, our measurement of \rhoHI\ is inconsistent with the $\mbox{\rhoHI}\propto (1+z)^{0.57}$ fit obtained by \citet{peroux-howk2020} --- which captures the evolution of \rhoHI\ with redshift from most DLA surveys to date --- at the 4$\sigma$ level (in linear space; note that Figure \ref{fig:rho_dla} is in logarithmic space). These remarkably low values for \rhoHI\ from Qz5 are also recovered when the Keck/ESI and VLT/X-SHOOTER samples are treated separately and independently (see Table \ref{table:simulated_crighton}). Furthermore, adopting the lower value of $\delta_{\scriptsize \HI} = 1.2$ would only result in a lower \rhoHI. We evaluate and discuss possible explanations for this deviation in Section \ref{5}.

\begin{deluxetable*}{ccc}
\tabletypesize{\normalsize}
\label{table:simulated_crighton}
\tablecaption{Measurements of $\ell_{\rm DLA}(X)$ and \rhoHI\ at $z\sim 5$ with different subsamples and at different spectral resolutions. Columns show the survey (1), the DLA incidence rate (2), and the cosmic \HI\ mass density at $z\sim 5$ (3).}
\setlength{\tabcolsep}{0.135in}
\tablehead{
\colhead{Sample} & \colhead{$\ell_{\rm DLA}(X)$} & \colhead{\rhoHI\,[$10^8\,\mbox{M}_{\odot}\,\mbox{Mpc}^{-3}$]}} 
\startdata 
Qz5 subDLAs (uncorrected for line blending) & - & $0.174_{0.143}^{0.245}$ \\
Qz5 subDLAs (corrected for $\approx 3\%$ line blending; \citealt{berg2019}) & - & $0.169_{0.139}^{0.238}$ \\
Qz5 DLAs (ESI only) & 0 & 0 \\
Qz5 DLAs (X-SHOOTER only) & $0.048_{0.03}^{0.07}$ & $0.57^{0.98}_{0.26}$ \\
Qz5 DLAs (full sample) & $0.034_{0.02}^{0.05}$ & $0.39^{0.68}_{0.18}$ \\
Qz5 DLAs (after matching the resolution and $S/N$ of \citealt{crighton2015}) & $0.094_{0.067}^{0.12}$ & $0.86^{1.37}_{0.53}$ \\
\citet{crighton2015} DLAs & $0.095_{0.07}^{0.12}$ & $1.2^{1.44}_{0.99}$ \\
\citet{crighton2015} & $0.095_{0.07}^{0.12}$ & $1.44^{1.73}_{1.19}$ \\
Qz5 DLAs (full sample) + subDLAs (corrected) & $0.034_{0.02}^{0.05}$ & $0.56^{0.82}_{0.31}$
\enddata
\end{deluxetable*}

Figure \ref{fig:rho_all} shows how \rhoHI\ varies with redshift after incorporating all techniques used for its estimation, with most measurements retrieved from the compilations by \citet{walter2020} and \citet{peroux-howk2020}. The techniques plotted include 21~cm emission (\citealt{zwaan2005,braun2012,jones2018}), 21~cm emission stacking (\citealt{lah2007,delhaize2013,rhee2013,hoppman2015,kanekar2016,bera2019,chowdhury2020}), 21~cm absorption (\citealt{grasha2020}), and Mg\,II absorption (\citealt{rao2017}). DLAs are plotted as black pentagons, and we note that we excluded all low spectral resolution measurements beyond $z\gtrsim 3.5$ (\citealt{noterdaeme2009,prochaska-wolfe2009,crighton2015,sanchez-ramirez2016}). All values plotted in Figures \ref{fig:rho_dla} and \ref{fig:rho_all} are reported in Appendix \ref{appendix3}.

All together, the redshift evolution of \rhoHI\ is similar to that of $\ell_{\rm DLA}(X)$, with an increase from the highest measured redshifts until $z\sim3$ and then steadily decreasing toward $z\sim0$. The $\mbox{\rhoHI}\propto (1+z)^{0.57}$ fit by \citet{peroux-howk2020} accurately characterizes this trend at $z<3$, but is in tension with our measurement at $z\sim 5$. Independently, some data points from studies at $z\sim4$ also suggest a lower \rhoHI\ at $z>3$ \citep{peroux2003,guimaraes2009,songaila-cowie2010,zafar2013}. As we will show in Section \ref{5.2}, the data are better reproduced by models where \rhoHI$(z)$ increases with cosmic time, peaks at $z\sim 3$, and then decreases at lower redshifts.

\section{Discussion} 
\label{5}
Here, we discuss several possible interpretations for the low $\ell_{\rm DLA}(X)$ and \rhoHI\ at $z\sim 5$ recovered by Qz5. We start by quantifying the impact that survey design --- primarily spectral resolution --- has on the measured values for $\ell_{\rm DLA}(X)$ and \rhoHI. We use this analysis to determine whether differences on the measured \rhoHI\ by different DLA surveys can be ascribed to systematic errors and/or statistical variations associated with the choice of sightlines. Then, we interpret the lower \rhoHI\ value at $z\sim 5$ within the context of our galaxy formation picture.

\subsection{On the measurements of \rhoHI\ at $z\sim 5$ from DLA surveys}
\label{5.1}

Several DLA surveys have attempted to measure \rhoHI\ over the last two decades. Some of the earliest constraints (\citealt{wolfe1995}) --- and those that followed (e.g. \citealt{prochaska-wolfe2009}) based on data from the SDSS Data Release 5 (\citealt{sdss-dr5}) --- are consistent with an increase in \rhoHI\ across the $z\sim 2-4.5$ range. However, the studies by \citet{rafelski2012,noterdaeme2012}, and \citet{neeleman2016} have pointed out that line blending in the \lya\ forest can lead to issues in the identification of DLAs at relatively low spectral resolutions. For data from the SDSS, which features a resolution of $R\sim 2000$, DLA identification is unreliable beyond $z\gtrsim 3.5$ (\citealt{noterdaeme2012}). Further complicating the issue is spectral S/N, which for SDSS can be prohibitively low at these redshifts (S/N~$\lesssim 3$). For these reasons, all DLA-based constraints on \rhoHI\ measured in SDSS data --- including the Baryon Oscillation Spectroscopic Survey of SDSS-III (BOSS; \citealt{dawson2013}) --- are highly uncertain (e.g. \citealt{prochaska-wolfe2009,bird2017,ho-bird-garnett2020}).

The limitations of SDSS data at these redshifts motivated \citet{crighton2015} to perform a dedicated search for DLAs at $z\sim 5$ at significantly higher S/N and slightly lower spectral resolution. We downloaded their data and measured a median S/N~$\sim 15$ in the \lya\ forest for their QSOs at $z>4.5$ using the same approach that we adopted for Qz5 in Table \ref{table:sample}, while their resolution is reported to be $R\sim 1300$ (\citealt{worseck2014}). They deemed their measurement of the DLA incidence rate uncertain, as it is strongly dependent on the number of DLAs with \NHI~$< 10^{20.5}$~cm$^{-2}$ that are challenging to accurately measure at low spectral resolutions. However, because \rhoHI\ is instead mostly driven by higher \NHI\ DLAs, they consider their estimate of \rhoHI\ to be more reliable. They recover a value of \rhoHI~$\approx1.44\times 10^8~\mbox{M}_{\odot}$~Mpc$^{-3}$, which is inconsistent with our measurement at the $\sim 4\sigma$ level.

The discrepancy in the value of \rhoHI\ at $z\sim 5$ between \citet{crighton2015} and our work may be due to differences in either spectral resolution and/or statistical sampling. Note that while the signal-to-noise per pixel of S/N~$\sim 15$ in \citet{crighton2015} is higher than ours (Table \ref{table:sample}), our signal-to-noise per \AA\ is effectively higher by $\approx 30\%$ if we account for the differences in spectral resolution and in spectral dispersion.

To study the impact of these effects on measurements of $\ell_{\rm DLA}(X)$ and \rhoHI, we downgraded the spectral resolution, lowered the spectral dispersion, and added noise to the Qz5 data to match that of the survey by \citet{crighton2015}. We adopted $R\sim 1300$, a pixel size of $0.46$~\AA, and a median S/N per pixel of~$\sim 15$ in the \lya\ forest, which are all representative of their final survey output (Giant Gemini GMOS survey; \citealt{worseck2014}). Then, the reproduced data were subject to the same DLA detection and analysis methodology described in Section \ref{3.2}, with two exceptions. For consistency with the approach followed by \citet{crighton2015}, we excluded all metal line information and opted for a single QSO continuum fit. We note that the method to search for DLAs adopted by \citet{crighton2015} was visual inspection, in similar fashion to our work.

\begin{figure}[t]
\centering
\includegraphics[width=3.35in]{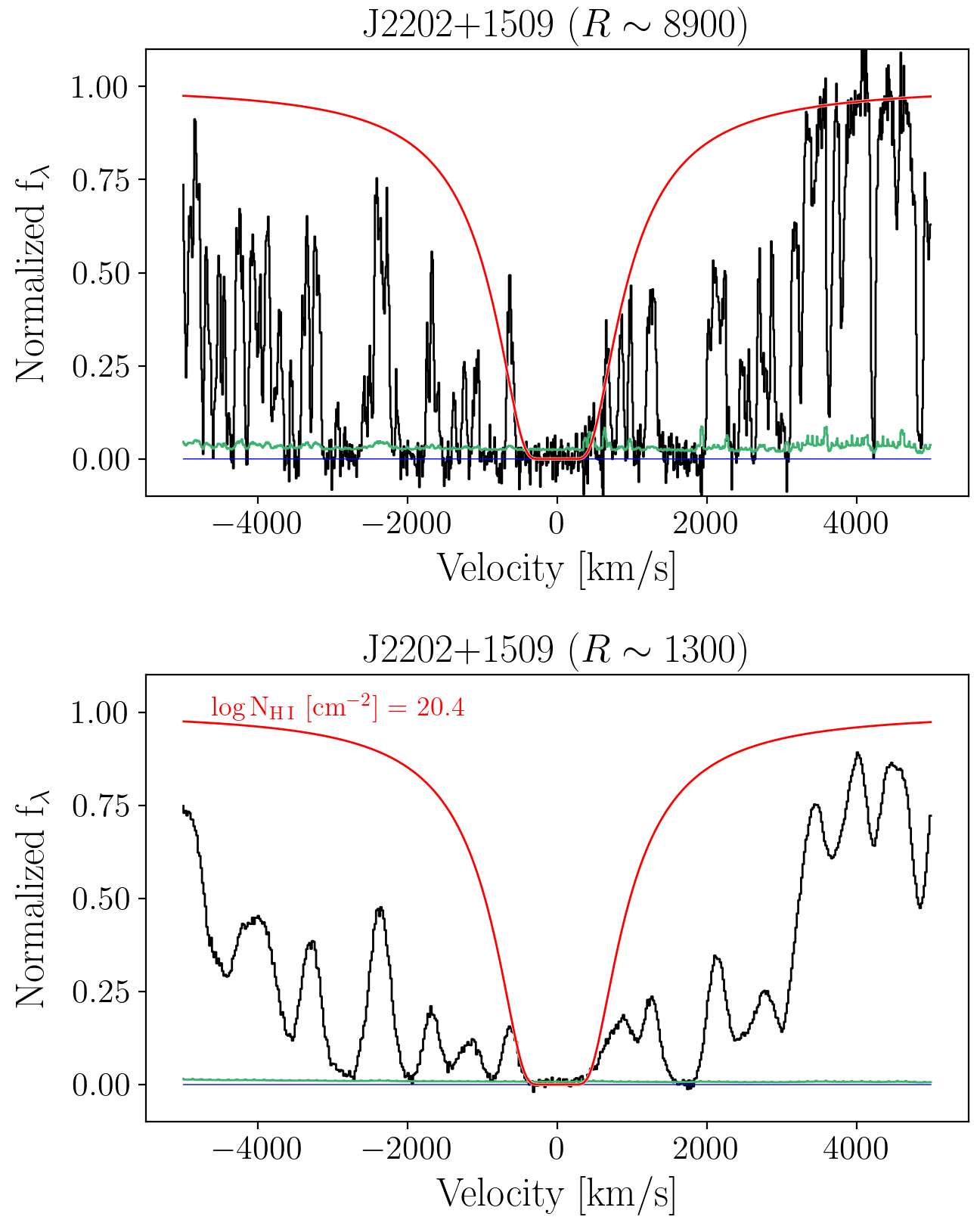}
    \caption{Example of how spectroscopy at low resolution can lead to false DLA detections at $z\gtrsim 3.5$. High resolution data is shown on top and simulated, lower resolution data for the same system is shown at the bottom. The spectra are plotted in black, the errors in green, and the zero-level fluxes in blue. The high-column density DLA that is apparent in the low resolution data (\NHI~$=10^{20.4}\mbox{cm}^{-2}$; red) is inconsistent with the high resolution spectrum. Mis-identifications of this kind can lead to overestimated values for \rhoHI\ at $z\gtrsim 3.5$ (see Table \ref{table:simulated_crighton}).}
   \label{fig:rho_systematics}
\end{figure}

We found a total of 14 DLAs in the downgraded spectra, which corresponds to almost three times the number of DLAs detected in the original, high resolution dataset (see Figure \ref{fig:rho_systematics} for an example). At first glance, this spurious DLA rate ($9/14\approx 64\%$) is at odds with the $\lesssim 15-30\%$ rate estimated by \citet{crighton2015} on high-resolution and/or in mock spectra. However, we should note that \citet{crighton2015} estimate that this number can reach $\approx 55\%$ at \NHI~$\sim10^{20.3}-10^{20.6}$~cm$^{-2}$, which is exactly the same \NHI\ regime in which we found most of the spurious DLAs.

We should note that we detected almost all originally identified DLAs (4/5) in the downgraded, low resolution spectra. This result is not in disagreement with \citet{crighton2015}, who determined that $\sim 80-90\%$ of DLAs are recovered at lower resolutions. Although differences between the $\log$~\NHI\ measured in the low and high resolution datasets can be significant (0.2~dex), we do not identify any systematic biases in the ability to recover the \NHI\ of real DLAs accurately. This was also the conclusion reached by \citet{crighton2015} from their simulations. 

Our measurements from the downgraded data were used to estimate a low resolution counterpart of \rhoHI\ from our survey, and the outcome is reported in Table \ref{table:simulated_crighton}. Comparison reveals that estimates of $\ell_{\rm DLA}(X)$ and \rhoHI\ from the downgraded Qz5 and the survey by \citet{crighton2015} are consistent within the uncertainties. We note that we are able to replicate their measurements both before and after they implement multiplicative resolution correction factors (since they are small; $\lesssim 10\%$). As apparent in Table \ref{table:simulated_crighton}, our analysis suggests that resolution correction factors are much larger ($\sim 100\%$ for \rhoHI).

It is evident from the procedures described above that lower spectral resolutions can lead to systematically higher recovered values for both $\ell_{\rm DLA}(X)$ and \rhoHI. For this reason, the \citet{crighton2015} data point at $z\sim 5$ in Figures \ref{fig:rho_dla} and \ref{fig:rho_all} is plotted as an upper limit and will not be considered in the analysis that follows in Section \ref{5.2}.

\subsection{Statistical significance of the low \rhoHI\ at $z\gtrsim 3.5$}
\label{5.2}

As apparent from the error bars in Figures \ref{fig:l_x} and \ref{fig:rho_dla}, our measurements of both $\ell_{\rm DLA}(X)$ and \rhoHI\ are not particularly precise. The uncertainties on $\ell_{\rm DLA}(X)$ and \rhoHI\ are dominated by the limited search path of Qz5, which features only 63 sightlines. This is because attaining large search paths at $z>3$ is difficult, since faint QSOs at these redshifts require massive programs at large telescopes. This can be appreciated in how the typical uncertainties on $\ell_{\rm DLA}(X)$ and \rhoHI\ in the literature increase toward $z>3$ in Figures \ref{fig:l_x} and \ref{fig:rho_dla}. 

Even so, other results from the literature suggest that low values for \rhoHI\ are not driven by the particular sightlines comprising Qz5. First, we have shown that the survey by \citet[which features a lower search path than ours at $z>5$]{crighton2015} is affected by resolution issues, and thus their value for \rhoHI\ at $z\sim 5$ is uncertain and likely overestimated. Second, the measurements of \rhoHI\ at $z>3$ by \citet{guimaraes2009,peroux2003,songaila-cowie2010}, and \citet{zafar2013} are consistent with a decrease in \rhoHI\ with redshift (see Figure \ref{fig:rho_dla}). To quantify this, we can compare two different \rhoHI\ models: one that is flat/increasing with redshift (Model $\mathcal{A}$) and another where \rhoHI\ is allowed to increase, peak, and then decrease with redshift (Model $\mathcal{B}$). We can make Model $\mathcal{A}$ match the parameterization suggested by \citet{peroux-howk2020}, which has the functional form
\begin{equation}
\dfrac{\mbox{\rhoHIa}(z)}{10^8~\mbox{M}_{\odot}~\mbox{Mpc}^{-3}} = a\,(1+z)^{\gamma}.
\end{equation}
For the second model, we considered the parameterization 
\begin{equation}
\dfrac{{\mbox{\rhoHIb}}(z)}{10^8~\mbox{M}_{\odot}~\mbox{Mpc}^{-3}} = \,c_0 + c_1\,(1+z) + c_2\,(1+z)^2.
\end{equation}

Both models were fitted to the data in Figure \ref{fig:rho_all} after careful assessment of redundant, correlated datapoints. We included the 21~cm emission measurements by \citet{lah2007,braun2012,delhaize2013,rhee2013,hoppman2015,jones2018,bera2019} and \citet{chowdhury2020}, while the constraints by \citet{zwaan2005} and \citet{kanekar2016} where excluded due to their respective redundancy with the measurements by \citet{delhaize2013} and \citet{chowdhury2020}. The estimates from 21~cm absorption (\citealt{grasha2020}) and from Mg\,II selected absorbers (\citealt{rao2017}) were also included. Finally, we utilized the datapoints from the DLA surveys by \citet{guimaraes2009,noterdaeme2012,zafar2013,neeleman2016,sanchez-ramirez2016}, and Qz5. Because of the inclusion of the study with SDSS by \citet{noterdaeme2012}, we excluded the measurements by \citet{noterdaeme2009} and \citet{prochaska-wolfe2009}. Similarly, because of the inclusion of the work by \citet{zafar2013}, we removed the constraints by \citet{peroux2003} and \citet{songaila-cowie2010}.

We decided for a Bayesian approach to constrain the posterior distributions of these models. We adopted flat, bounded priors for all parameters, and we fitted for asymmetric errors in the \rhoHI\ estimates. The posteriors were quantified in multidimensional grids for all model parameters.

To determine what model best reproduces the data, we can turn to the Bayes factor. This quantity is defined as
\begin{equation}
B_{\mathcal{AB}} = p_{\mathcal{A}}/p_{\mathcal{B}}, 
\end{equation}

where $p_{\mathcal{A}}$ and $p_{\mathcal{B}}$ correspond to the marginal likelihoods of models $\mathcal{A}$ and 
$\mathcal{B}$, respectively. Note that the use of the marginal likelihoods makes the Bayes factor ideal for comparing models with different number of parameters (i.e., Model $\mathcal{B}$ is penalized for having one more free parameter than Model $\mathcal{A}$). We obtain a value of 
\begin{equation}
\log{B_{\mathcal{AB}}} < -5,
\end{equation}
which according to Jeffrey's scale corresponds to decisive evidence in favor of Model $\mathcal{B}$. Thus, we conclude that literature measurements of \rhoHI\ are most consistent with a model that allows for \rhoHI$(z)$ to increase with cosmic time, peak, and then decrease toward low redshift. 

We found the best-fit parameters of Model $\mathcal{B}$ to be
\begin{equation}
c_0 = -0.31 \pm 0.02,
\end{equation}
\begin{equation}
c_1 = 0.8 \pm 0.02, \mbox{ and}
\end{equation}
\begin{equation}
c_2 = -0.1035 \pm 0.0035.
\end{equation}

The solution achieved with these parameters is plotted in Figure \ref{fig:rho_all}.

\begin{figure*}
\centering
\includegraphics[width=7.1in]{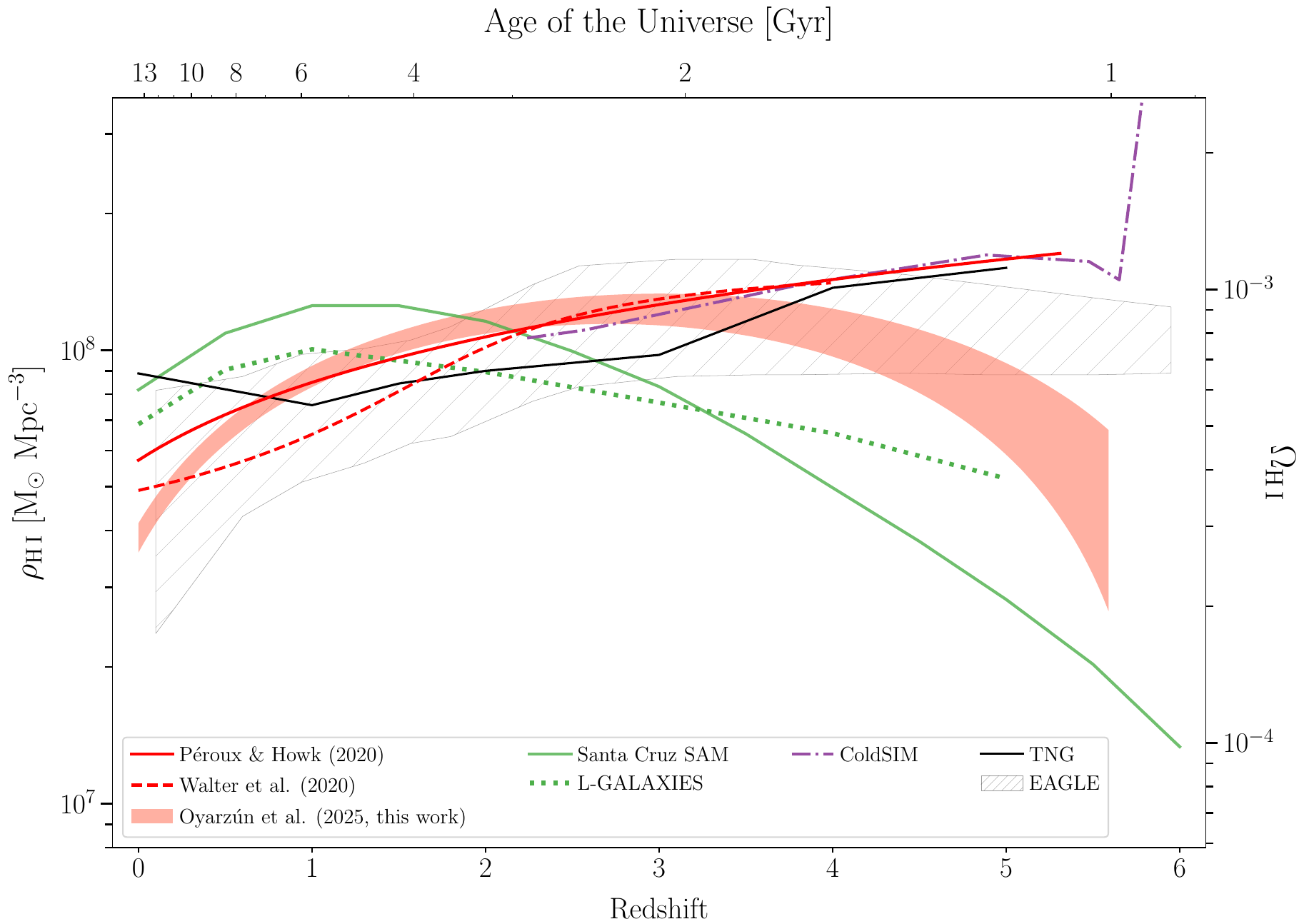}
    \caption{The cosmic \HI\ mass density as a function of redshift. Plotted are constraints from the data (\citealt{peroux-howk2020,walter2020} and this work), semi-analytic models (Santa Cruz and L-GALAXIES; \citealt{popping2014,henriques2020}), hydrodynamical simulations with an explicit \HI\ component (ColdSim; \citealt{maio2022}), and hydrodynamical simulations without built-in partitioning of the different hydrogen components (IllustrisTNG and EAGLE; \citealt{vogelsberger2014a,rahmati2015}).}
   \label{fig:rho_sims}
\end{figure*}

\subsection{Interpretation for the low \rhoHI\ at $z\gtrsim 3.5$}
\label{5.3}

We have concluded that observations show evidence for an increase in \rhoHI\ from $z \sim 5$ to $z\sim 3$. To interpret this trend, we can turn to analytic arguments for the evolution of the baryonic components of galaxies with cosmic time (e.g. \citealt{walter2020}). In particular, we are interested in the mechanisms that dictate the assembly of \HI\ gas in galaxies.

The baryonic budget of galaxies is formed and replenished through the accretion (and cooling) of neutral and ionized gas from the Cosmic Web (e.g. \citealt{bouche2010,tacconi2018}). Thus, the rate at which the baryonic budgets of galaxies grow with cosmic time is tied to the rate at which halos accrete matter. Taking the halo mass accretion rate in $\Lambda$CDM (\citealt{rodriguez-puebla2016}) and a baryonic matter fraction that is constant with redshift yields baryonic mass densities inside galaxies ($\rho_{\rm b, gal}$) that increase monotonically with cosmic time (\citealt{walter2020}). This behavior is a direct result of how the halo mass function grows as the Universe evolves.

Some theoretical models prescribe an evolution with redshift for \rhoHI\ that is very similar to that of $\rho_{\rm b, gal}$. In these prescriptions, the increase of \rhoHI\ with time naturally follows accretion of matter from the cosmic web. In a two step-process, this accretion of gas is turned into stars through (1) the infall of neutral and ionized onto extended \HI\ gas reservoirs through cold mode accretion and (2) the cooling of this gas that leads to the production of H$_2$ and the formation of stars \citep[e.g.][]{walter2020}. Therefore, the increase in \rhoHI\ with cosmic time is followed by an increase in the cosmic molecular gas mass density ($\rho_{\tiny \mbox{H}_2}$) and an increase in the cosmic star-formation rate density (\SFRD). 

It stands to reason that the normalization of \rhoHI$(z)$ is also directly linked to how and when \HI\ mass builds up in galaxies. For instance, \citet{popping2014} and \citet{berry2014,berry2016} --- who utilized the semi-analytic models developed by \citet{somerville-primack1999} and \citet{somerville2008,somerville2012} to study the evolution of \rhoHI\ with cosmic time --- predict \rhoHI\ to increase with time before turning over and decreasing toward $z<1$ (see Figure \ref{fig:rho_sims}). Similar trends were found in the \texttt{GALFORM} (\citealt{lagos2014}), \texttt{GAEA} (\citealt{spinelli2020}), and \citet{theuns2021} semi-analytic models, as well as in the L-GALAXIES simulation (\citealt{martindale2017,henriques2020}). On the other hand, the build-up of \HI\ occurs mostly at $z>4$ in other studies (Figure \ref{fig:rho_sims}). In the EAGLE simulation (\citealt{rahmati2015}), \rhoHI\ only starts to show evidence of an initial increase toward $z\gtrsim 3.5$. In IllustrisTNG (\citealt{vogelsberger2014a}) and in ColdSim (\citealt{maio2022}), \rhoHI\ builds up before $z\sim 5$. Here, the IllustrisTNG constraint corresponds to the average between the measurements by \citet{villaescusa-navarro2018} and \citet{popping2019} (as reported in Figure 3 of \citealt{yates2021}). Our analysis of the observational data reveals that this build-up is still ongoing at $z\sim 5$ and continues out to at least $z\sim 4$.

As pointed out by \citet{yates2021}, differences in the amount and/or the column density of \HI\ gas in the CGM of galaxies between theoretical work can sometimes exceed $\gtrsim 50\%$. Thus, the presence (or lack thereof) of high column density \HI\ gas far out into the IGM and/or the CGM of galaxies (beyond 2 effective radii; \citealt{byrohl2021,yates2021}) also emerges as an important factor in determining the normalization of \rhoHI. By extension, variations in the amount of \HI\ in the CGM of galaxies with cosmic time could also impact the evolution of \rhoHI$(z)$.

Evidence suggests that \rhoHI\ (as derived by DLA surveys) is indeed sensitive to the amount of \HI\ gas in the CGM. First,  DLAs typically contain little H$_2$ (based on the lack of Lyman-Werner absorption; e.g. \citealt{jorgenson2014}), hinting that DLAs rarely trace gas near star-forming regions (as predicted by simulations; e.g. \citealt{stern2021,stern2023}). Second, the projected separations between DLAs and associated galaxies can exceed $40$~kpc at $z\sim 2-4$, which is indicative of CGM scales (e.g. \citealt{fumagalli2017,krogager2017,neeleman2017,neeleman2018,neeleman2019,mackenzie2019,prochaska2019,kanekar2020,rhodin2021,kaur2021,kaur2022a,kaur2022b,lofthouse2023,oyarzun2024}; see also \citealt{christensen2014,hamanowicz2020}). This is consistent with the declining incidence rate of \MgII\ absorbers toward $z>2$ (\citealt{matejek-simcoe2012}), under the assumption that these \MgII\ trace the population of DLAs at these redshifts (e.g. \citealt{berg2017,rao2017}). As samples of \HI-selected galaxies continue to grow, it will be interesting to establish if any evolution with redshift is observed in the projected separations between galaxies and DLAs. 

In this framework that envisions \rhoHI\ evolving with $\rho_{\rm b, gal}$, one would not expect that \rhoHI\ before the formation of the first stars corresponds to the cosmic mass density of baryonic matter. As redshift increases, we expect the fraction of \rhoHI\ that is present in low \NHI\ gas in the IGM to increase, driving constraints on \rhoHI\ --- as measured by probes of high \NHI\ gas --- apart from the true value of \rhoHI. We identify tentative evidence that this distinction becomes relevant at high redshift in the  major contribution of subDLAs to \rhoHI. While we know that subDLAs probe $10-20\%$ of the \HI\ gas mass in the Universe at $z\sim 2-4$ (e.g. \citealt{zafar2013,berg2019}), we recover a higher value of $\approx 30\%$ at $z\sim 5$ with Qz5 (albeit with limited statistical power). In summary, and as also pointed out by \citet{heintz2022}, there is a distinction to be made between \rhoHI\ (as probed by gas in the ISM and CGM of galaxies) and true \rhoHI\ (which also accounts for gas in the IGM), especially at very high redshift.

\newpage

\subsection{Alternative scenarios for the low \rhoHI\ at $z\gtrsim 3.5$}
\label{5.4}

An alternative scenario for the lower value of \rhoHI\ at $z\gtrsim 3.5$ may emerge from recent studies of the \lya\ and Ly$\beta$ forests, where fluctuations in the neutral gas fraction down to $z\sim 5-6$ have been identified (\citealt{becker2015,bosman2018,bosman2022,qin2021,grazian2023,zhu2023,zhu2024,spina2024}). In this context, one may consider that the end of the reionzation of the Universe is somehow connected with the changes in \rhoHI\ that we observe. Evidence suggests that a direct connection is unlikely, with recent studies revealing that high \NHI\ gas can form well before the end of reionization (DLAs at $z\sim 10$; \citealt{heintz2024}). Thus, our current understanding is that the formation of high column density \HI\ gas and the reionization of the intergalactic medium are not necessarily coupled and that they can occur at widely different redshifts.

It is also worth discussing whether the turnover in \rhoHI\ at $z\gtrsim 3.5$ could be driven by redshift variations in the sensitivity of absorption studies to high \NHI\ gas. Dust extinction stands out as a the primary candidate for this potential bias, with studies dating back more than three decades showing that QSO sightlines containing DLAs feature higher dust reddening than QSO sightlines lacking in DLAs (\citealt{fall1989,fall1996,pei1991,pei1999,fall1993,pei1995}). That said, studies with SDSS and DESI have revealed that this effect is minor, at least out to $z\sim 4$ (\citealt{murphy2016,napolitano2024}). If this effect was to become more significant toward $z\gtrsim 4$, it could contribute to and/or fabricate the turnover in \rhoHI\ that we report in this paper. 

\section{Summary} 
\label{6}

We have conducted Qz5, a spectroscopic survey of 63 bright QSOs at $z\sim 4.7-5.7$ with Keck/ESI and VLT/X-SHOOTER that is aimed at constraining $\ell_{\rm DLA}(X)$ and \rhoHI\ at $z\sim 5$. Qz5 stands out because of the high resolution ($R\sim7000-9000$) of the spectra, enabling us to constrain $\ell_{\rm DLA}(X)$ and \rhoHI\ with accuracy by minimizing biases that that can affect studies at low spectral resolutions. We found a total of 5 non-proximate DLAs in the 63 QSOs, which translates into a DLA incidence rate of $\ell_{\rm DLA}(X) \sim 0.034_{0.02}^{0.05}$. We estimate that the contribution of subDLAs toward \rhoHI\ is higher at $z\sim 5$ ($\approx30\%$) than at lower redshifts ($\approx17\%$). Accounting for the contribution of both DLAs and subDLAs, we measure a value for the cosmic \HI\ mass density at $z\sim 5$ of \rhoHI~$=0.56^{0.82}_{0.31}\times 10^8~\mbox{M}_{\odot}$~Mpc$^{-3}$. These values for $\ell_{\rm DLA}(X)$ and \rhoHI\ are $\sim 2-4\sigma$ lower than previously reported values at $z\sim 2-5$.

To understand the discrepancy between our measurements and the results from previous work, we studied the impact of spectral resolution on estimates of $\ell_{\rm DLA}(X)$ and \rhoHI. We focused on replicating the observational setup by \citet{crighton2015}, which stands out as a high S/N study of \rhoHI\ at $z\sim 5$ from DLAs (S/N~$\sim 15$, $R\sim 1300$). We show that lowering the resolution of our survey --- from $R \gtrsim 7000$ to $R\sim 1300$ --- leads to systematically higher estimates of $\ell_{\rm DLA}(X)$ and \rhoHI, accounting for any discrepancies. We therefore conclude that high resolution spectroscopy is critical for accurately measuring $\ell_{\rm DLA}(X)$ and \rhoHI\ toward $z\gtrsim 3.5$.

Our measurements at $z\sim 5$ indicate that $\ell_{\rm DLA}(X)$ and \rhoHI\ at $z\sim 5$ are lower than the widely adopted fit to the observational data by \citet{peroux-howk2020} with 4$\sigma$ significance. After the exclusion of biased constraints at $z>3.5$ and the inclusion of our new measurement at $z\sim 5$, the preference for a \rhoHI = \rhoHI$(z)$ model that peaks at $z\sim 3$ is decisive, according to a Bayesian comparison. This means that observations of \rhoHI$(z)$ from DLAs are best reproduced by models that increase with cosmic time, peak at $z\sim3$, and then decrease at low redshift. 

In conclusion, we show for the first time that \rhoHI\ follows a similar evolutionary shape as the \SFRD\ and $\rho_{\tiny \mbox{H}_2}$, corresponding to a rise at early times from the inflow of gas from the Cosmic Web onto the halos of galaxies. This behavior is then followed by a peak at $z\sim3$ and a turnover at $z<2$. To constrain the highly uncertain peak of \rhoHI\ in cosmic time, additional high resolution, high S/N  surveys of QSOs at $z=3-5.5$ are needed. Regardless, it appears that \rhoHI\ peaks earlier than the \SFRD\ and $\rho_{\tiny \mbox{H}_2}$, which already constrains galaxy formation models.

We are thankful to the reviewer of this manuscript for the constructive comments and suggestions. The authors would also like to acknowledge Nissim Kanekar for the valuable conversations and insightful comments in regards to this manuscript. This material is based upon work supported by the National Science Foundation under grant No. 2107989. MR also acknowledges support from the STScI's Director's Discretionary Research Funding (grant ID D0101.90228). LC is supported by DFF/Independent Research Fund Denmark, grant-ID 2032–00071. SL acknowledges support by FONDECYT grant 1231187. This work made use of Astropy\footnote{\url{http://www.astropy.org}}, a community-developed core Python package and an ecosystem of tools and resources for astronomy \citep{astropy:2013, astropy:2018, astropy:2022}. This work also made use of \texttt{linetools} (\citealt{linetools}).

\clearpage

\appendix

\section{Q\lowercase{z}5 spectra}
\label{appendix1}

\begin{figure*}[b]
\centering
\includegraphics[width=7.1in]{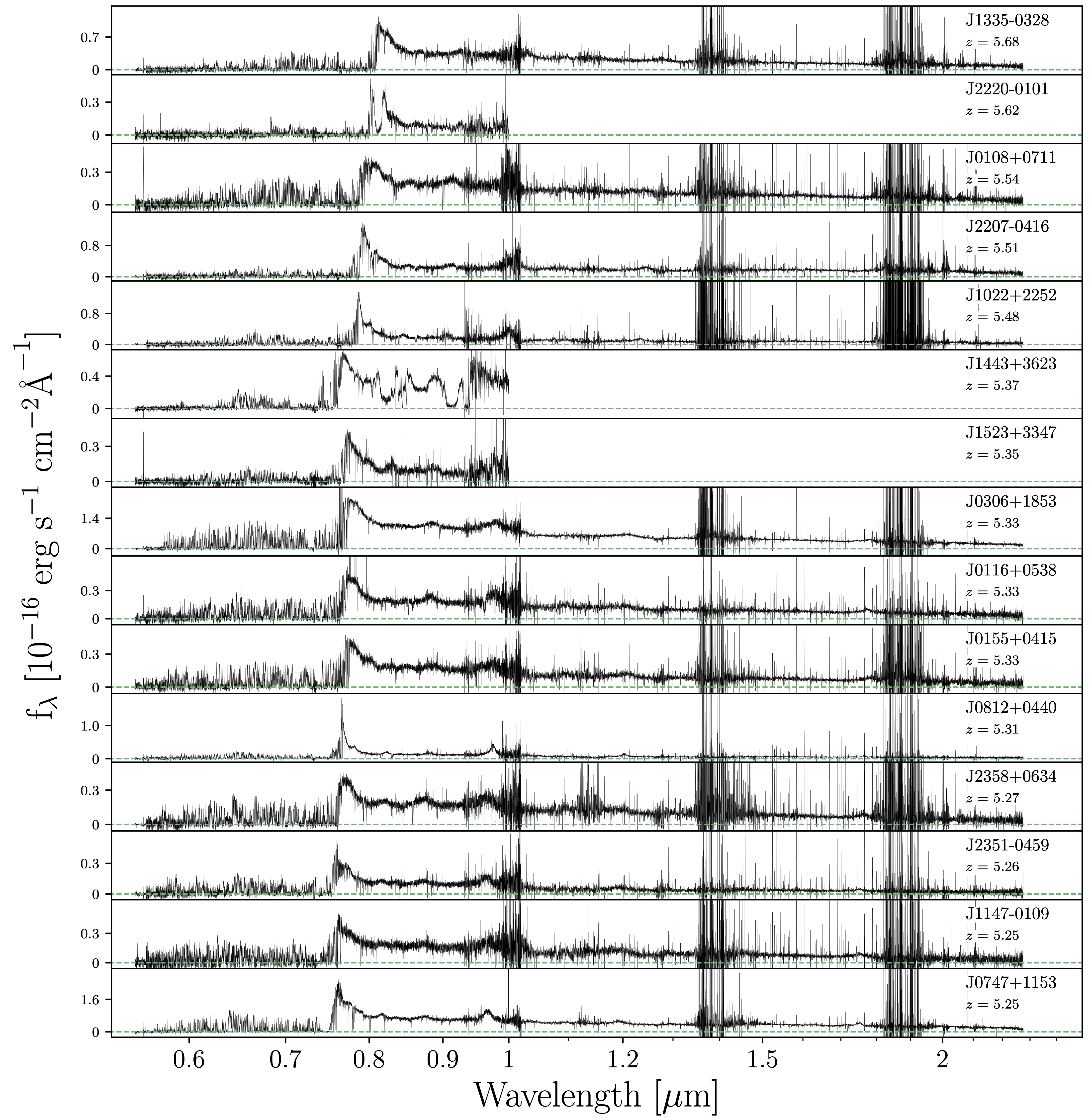}
\caption{}
\label{figA}
\end{figure*}

\begin{figure*}
\centering
\includegraphics[width=7.1in]{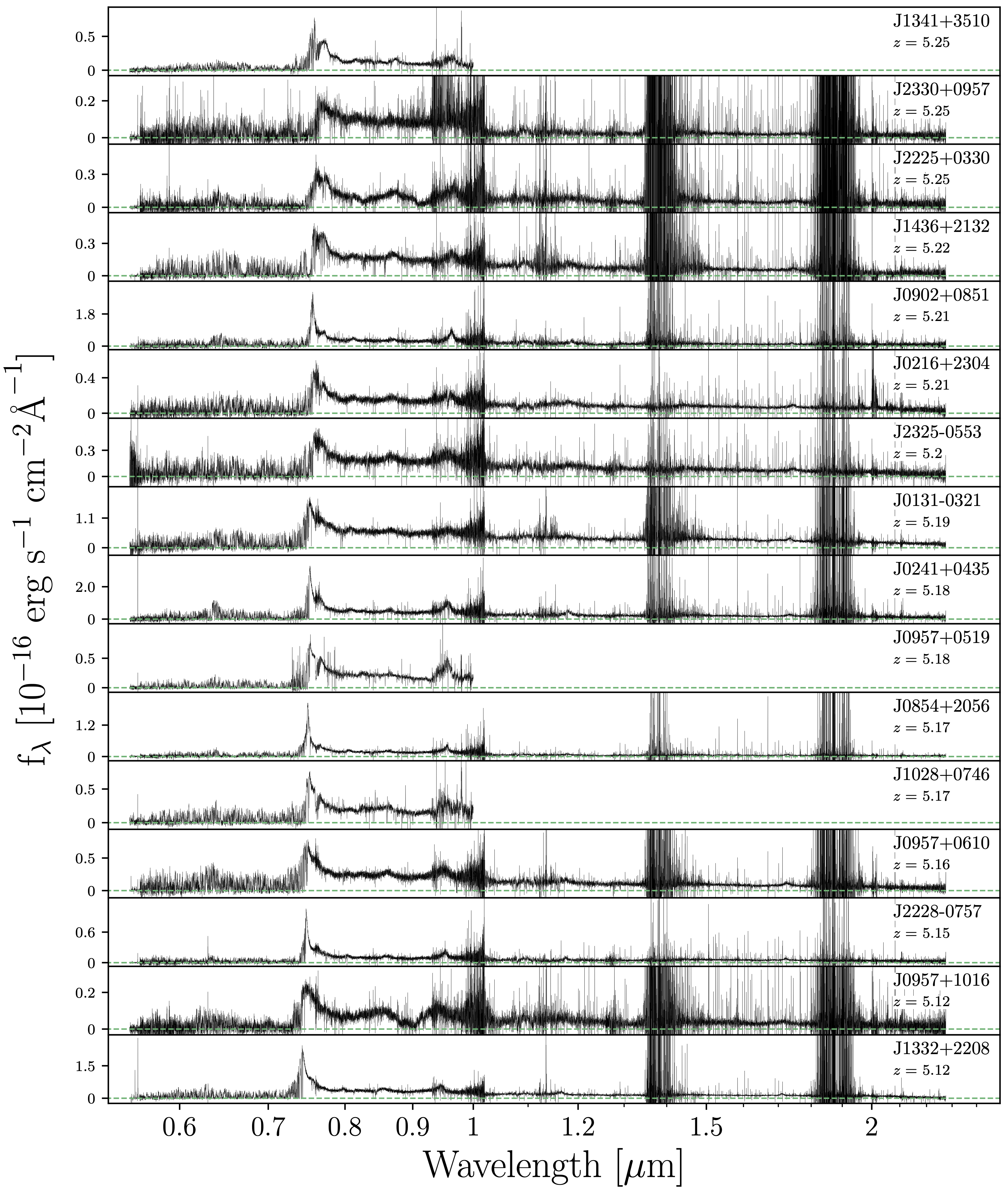}
\caption{}
\label{figB}
\end{figure*}

\begin{figure*}
\centering
\includegraphics[width=7.1in]{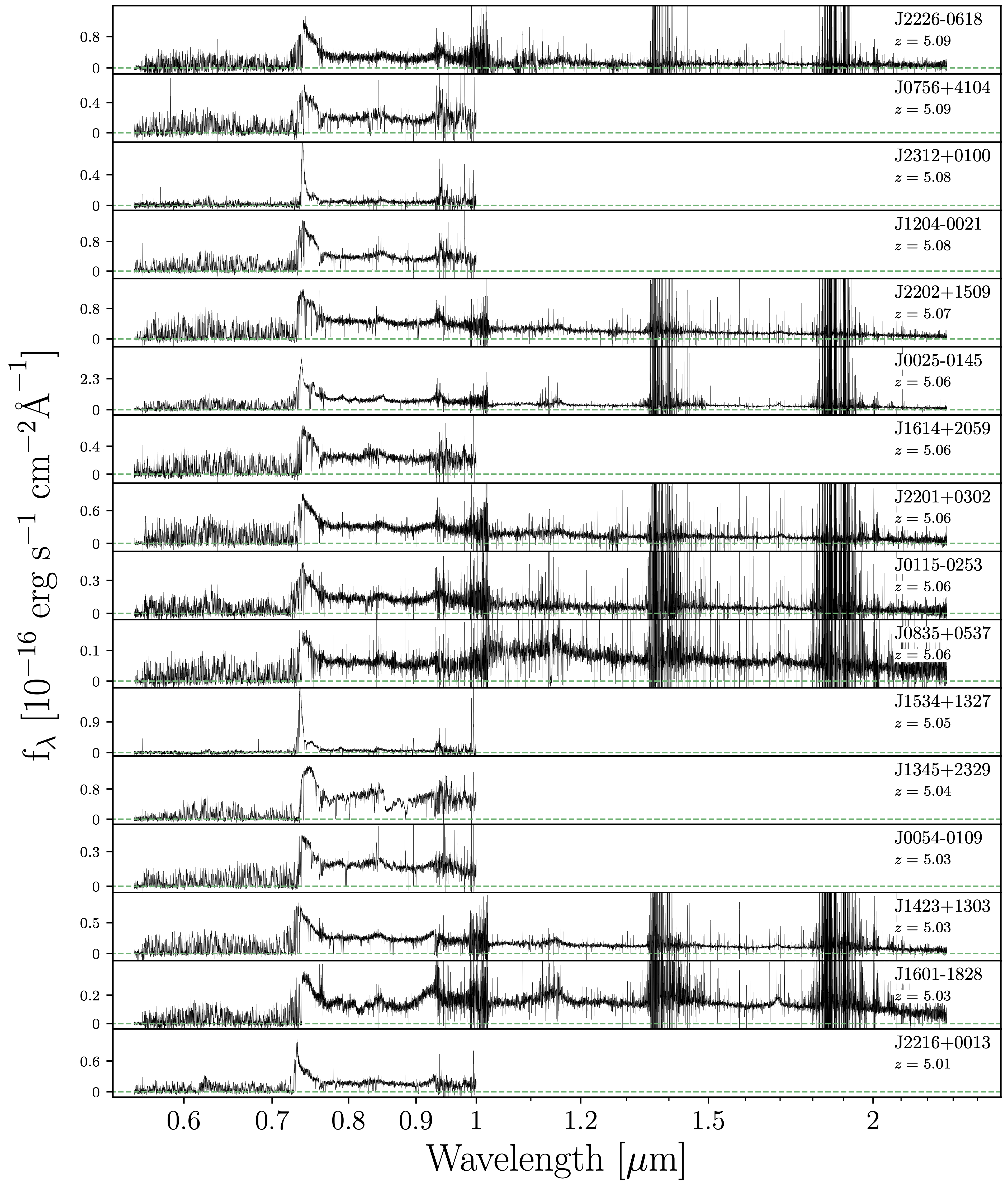}
\caption{}
\label{figC}
\end{figure*}

\begin{figure*}
\centering
\includegraphics[width=7.1in]{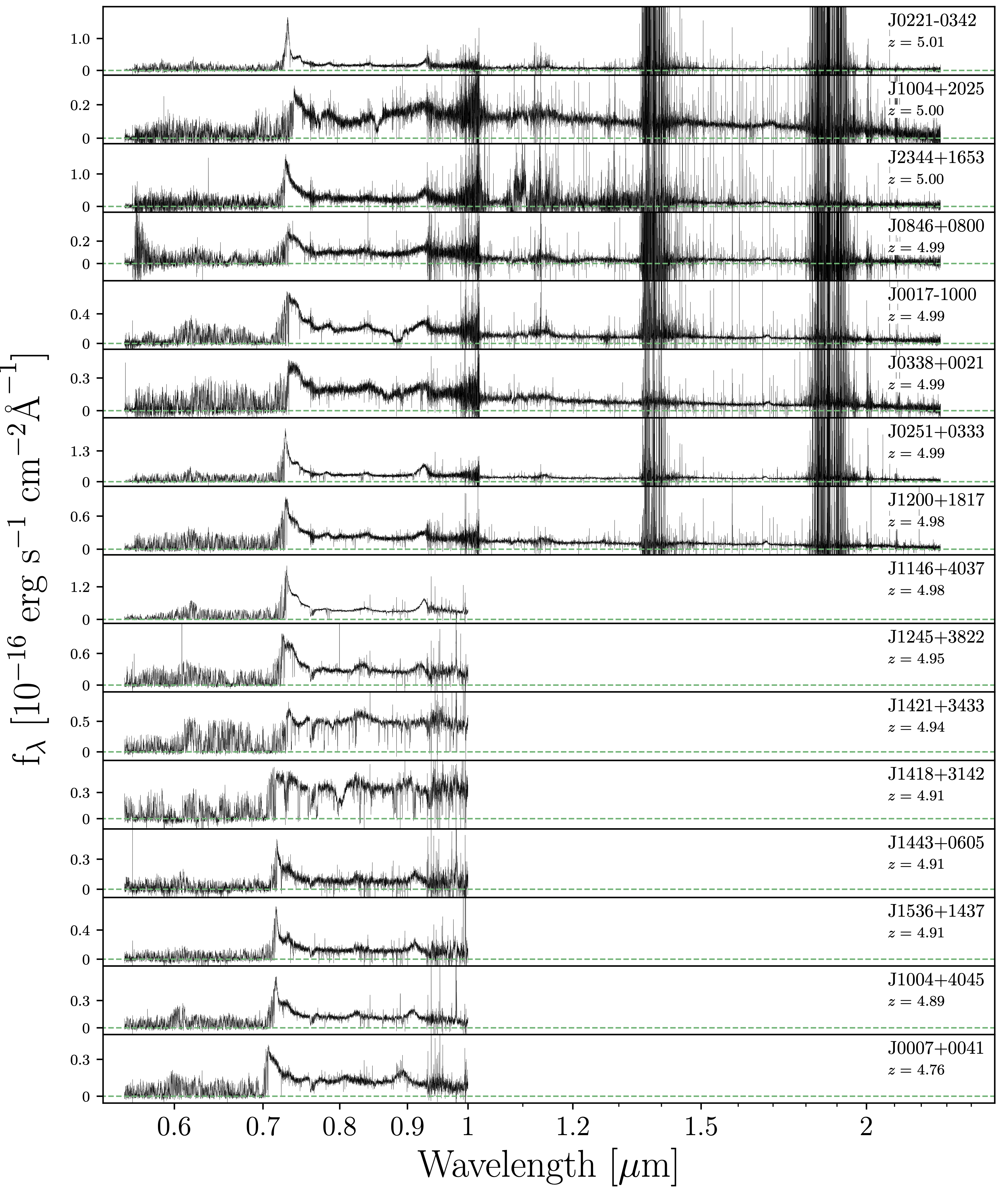}
\caption{}
\label{figD}
\end{figure*}

The data for the Qz5 survey is available for download at \url{https://doi.org/10.5281/zenodo.14825981} and \url{http://www.rafelski.com/data/DLA/qz5/}. The spectra are shown in Figures \ref{figA} through \ref{figD}. 

\newpage

\section{\HI\ absorbers identified in the spectra}
\label{appendix2}
We outline in Table \ref{table:appendix2} all \HI\ absorbers that were identified as DLA candidates during our search outlined in Section \ref{3.2}. The absorbers are grouped by QSO sightline and ordered according to the absorption redshift. The notes column specifies the reason for the inclusion or exclusion of the absorber into our analysis of $\ell_{\rm DLA}(X)$ and \rhoHI.

\startlongtable
\begin{deluxetable*}{ccccccc}
\tabletypesize{\normalsize}
\label{table:appendix2}
\setlength{\tabcolsep}{0.15in}
\tablehead{
\colhead{QSO} & \colhead{ra} & \colhead{dec} & \colhead{z$_{em}$} & \colhead{z$_{abs}$} & \colhead{Notes} &  \colhead{log\NHI~[cm$^{-2}$]}} 
\startdata 
J0007+0041 & 00:07:49.17 & 00:41:19.62 & 4.76  & 4.7330 & \xmark: Proximate DLA & 20.6 $\pm$ 0.3\\
J0025-0145 & 00:25:26.84 & -01:45:32.50 & 5.06 & 4.7389 & \cmark: DLA & 20.3 $\pm$ 0.15\\
J0116+0538 & 01:16:14.31 & 05:38:17.59 & 5.33 & 5.108 & \cmark: subDLA & 19 $\pm$ 0.2\\
... & ... & ... & ... & 5.113 & \cmark: subDLA & 19.3 $\pm$ 0.2\\
J0131-0321 & 01:31:27.35 & -03:21:00.08 & 5.19 & 4.654 & \cmark: subDLA & 19.0 $\pm$ 0.2\\
... & ... & ... & ... & 4.962 & \cmark: subDLA & 19.5 $\pm$ 0.2\\
J0306+1853 & 03:06:42.51 & 18:53:15.82 & 5.33 & 4.9866 & \cmark: DLA & 20.9 $\pm$ 0.15 \\
... & ... & ... & ... & 5.012 & \cmark: subDLA & 20.2 $\pm$ 0.2\\
J0747+1153 & 07:47:49.18 & 11:53:52.44 & 5.25 & 4.619 & \cmark: subDLA & 20.1 $\pm$ 0.2\\
... & ... & ... & ... & 5.1447 & \cmark: DLA & 21.1 $\pm$ 0.25\\
J0812+0440 & 08:12:48.82 & 04:40:56.57 & 5.31 & 5.108 & \cmark: subDLA & 20 $\pm$ 0.15\\
J0902+0851 & 09:02:45.76 & 08:51:15.90 & 5.21 & 4.586 & \cmark: subDLA & 19.3 $\pm$ 0.2\\
... & ... & ... & ... & 4.864 & \cmark: subDLA & 19.6 $\pm$ 0.2\\
... & ... & ... & ... & 5.070 & \cmark: subDLA  & 19.8 $\pm$ 0.2\\
J1004+2025 & 10:04:44.31 & 20:25:20.03 & 5.00 & 4.854 & \cmark: subDLA  & 20 $\pm$ 0.2\\
J1146+4037 & 11:46:57.79 & 40:37:08.59 & 4.98 & 4.533 & \cmark: subDLA  & 19 $\pm$ 0.2\\
... & ... & ... & ... &  4.748 & \cmark: subDLA  & 19.3 $\pm$ 0.1\\
J1147-0109 & 11:47:06.42 & -01:09:58.37 & 5.25 & 5.062 & \cmark: subDLA  & 19.6 $\pm$ 0.2\\
J1335-0328 & 13:35:56.24 & -03:28:38.29 & 5.68 & 4.647 & \cmark: subDLA & 19.9 $\pm$ 0.2\\
... & ... & ... & ... &  5.295 & \cmark: subDLA  & 19.7 $\pm$ 0.35\\
... & ... & ... & ... & 5.352 & \cmark: subDLA  & 20.1 $\pm$ 0.2\\
J1345+2329 & 13:45:26.62 & 23:29:49.30 & 5.04 & 5.0060 & \xmark: Proximate DLA & 21.1 $\pm$ 0.1\\
J1421+3433 & 14:21:03.83 & 34:33:32.00 & 4.94 & 4.665 & \cmark: subDLA & 20.2 $\pm$ 0.15\\
J1436+2132 & 14:36:05.00 & 21:32:39.26 & 5.22 & 5.1780 & \xmark: Proximate DLA & 20.7 $\pm$ 0.25 \\
J1443+3623 & 14:43:50.67 & 36:23:15.18 & 5.37 & 4.868 & \cmark: subDLA  & 19.9 $\pm$ 0.2\\
J1523+3347 & 15:23:45.69 & 33:47:59.41 & 5.35 & 4.797 & \cmark: subDLA  & 19.6 $\pm$ 0.2\\
J1534+1327 & 15:34:59.76 & 13:27:01.43 & 5.05 & 4.501 & \cmark: subDLA  & 19.5 $\pm$ 0.3\\
... & ... & ... & ... &  4.961 & \xmark: Proximate subDLA  & 19.9 $\pm$ 0.3\\
J1601-1828 & 16:01:11.17 & -18:28:35.07 & 5.03 & 5.052 & \xmark: Proximate subDLA  & 19.0 $\pm$ 0.1\\
J2202+1509 & 22:02:26.77 & 15:09:52.37 & 5.07 & 4.948 & \cmark: subDLA  & 19.7 $\pm$ 0.15\\
J2207-0416 & 22:07:10.13 & -04:16:56.22 & 5.51 & 4.7220 & \cmark: DLA & 20.4 $\pm$ 0.2\\
... & ... & ... & ... & 4.943 & \cmark: subDLA & 19.8 $\pm$ 0.2\\
... & ... & ... & ... & 5.036 & \cmark: subDLA  & 20.0 $\pm$ 0.2\\
... & ... & ... & ... & 5.141 & \cmark: subDLA  & 19.5 $\pm$ 0.3\\
... & ... & ... & ... & 5.3374 & \cmark: DLA  & 20.8 $\pm$ 0.15 \\
J2226-0618 & 22:26:12.42 & -06:18:07.35 & 5.09 & 4.726 & \cmark: subDLA & 19.7 $\pm$ 0.2\\
J2325-0553 & 23:25:36.64 & -05:53:28.42 & 5.20 & 4.940 & \cmark: subDLA  & 20.1 $\pm$ 0.2\\
J2351-0459 & 23:51:24.31 & -04:59:07.30 & 5.26 & 5.063 & \cmark: subDLA  & 20.0 $\pm$ 0.2\\
J2358+0634 & 23:58:24.05 & 06:34:37.48 & 5.27 & 4.905 & \cmark: subDLA  & 19.7 $\pm$ 0.2\\
\enddata
\end{deluxetable*}

\section{\rhoHI\ values}
\label{appendix3}

The observational constraints on \rhoHI\ used throughout this paper are reported in Table \ref{table:appendix3}. Values were adapted to $H_0 = 70~$km~s$^{-1}$~Mpc$^{-1}$. We also assume $\delta_{\scriptsize \HI} = 1.2$ for all DLA measurements, except for Qz5 ($\delta_{\scriptsize \HI} = 1.44$). Any corrections for the contribution from Helium (e.g. \citealt{peroux-howk2020}) were reversed.

\startlongtable
\begin{deluxetable*}{ccccccc}
\tabletypesize{\normalsize}
\label{table:appendix3}
\setlength{\tabcolsep}{0.04in}
\tablehead{
\colhead{$z$} & \colhead{$z_{bin}$} & \colhead{\rhoHI$\,\times\,10^{-8}$} & \colhead{\rhoHI$\,\times\,10^{-8}~(1\sigma)$} & \colhead{Original} & \colhead{Method} & \colhead{Retrieved} \\
\colhead{} & \colhead{} & \colhead{[M$_{\odot}$~Mpc$^{-3}$]} & \colhead{[M$_{\odot}$~Mpc$^{-3}$]} & \colhead{reference} & \colhead{} & \colhead{from}} 
\startdata 
0 & (0, 0) & 0.46 & (0.38, 0.54) & \citet{zwaan2005} & 21~cm & \citet{walter2020} \\
0 & (0, 0) & 0.79 & (0.66, 0.92) & \citet{braun2012} & 21~cm & \citet{walter2020} \\
0.03 & (0, 0.06) & 0.39 & (0.32, 0.46) & \citet{jones2018} & 21~cm & \citet{walter2020} \\
0.03 & (0, 0.04) & 0.50 & (0.45, 0.60) & \citet{delhaize2013} & 21~cm & \citet{walter2020} \\
0.1 & (0.04, 0.13) & 0.56 & (0.48, 0.66) & \citet{delhaize2013} & 21~cm & \citet{walter2020} \\
0.07 & (0, 0.12) & 0.28 & (0.20, 0.37) & \citet{hoppman2015} & 21~cm & \citet{walter2020} \\
0.1 & (0.08, 0.12) & 0.41 & (0.35, 0.47) & \citet{rhee2013} & 21~cm & \citet{walter2020} \\
0.2 & (0.16, 0.22) & 0.42 & (0.32, 0.53) & \citet{rhee2013} & 21~cm & \citet{walter2020} \\
0.24 & (0.24, 0.24) & 0.85 & (0.48, 1.23) & \citet{lah2007} & 21~cm & \citet{walter2020} \\
0.34 & (0.2, 0.4) & 0.48 & (0.41, 0.55) & \citet{bera2019} & 21~cm & \citet{walter2020} \\
1.06 & (0.74, 1.45) & 0.61 & (0.46, 0.76) & \citet{chowdhury2020} & 21~cm & \citet{chowdhury2020} \\
1.27 & (1.18, 1.34) & $<0.27^{*}$ & (0.00, 0.09) & \citet{kanekar2016} & 21~cm & \citet{walter2020} \\
0.41 & (0, 0.69) & 0.50 & (0.32, 0.67) & \citet{grasha2020} & 21~cm abs. & \citet{grasha2020} \\
1.1 & (0.69, 2.74) & 1.61 & (0.81, 2.42) & \citet{grasha2020} & 21~cm abs. & \citet{grasha2020} \\
0.46 & (0.11, 0.61) & 1.02 & (0.68, 1.37) & \citet{rao2017} & Mg~II & \citet{walter2020} \\
0.73 & (0.61, 0.89) & 0.89 & (0.67, 1.10) & \citet{rao2017} & Mg~II & \citet{walter2020} \\
1.17 & (0.89, 1.65) & 1.00 & (0.61, 1.38) & \citet{rao2017} & Mg~II & \citet{walter2020} \\
0.62 & (0.01, 1.6) & 0.30 & (0.16, 0.54) & \citet{neeleman2016} & DLAs & \citet{walter2020} \\
1.84 & (1.51, 2) & 1.05 & (0.81, 1.28) & \citet{zafar2013} & DLAs$^{\dagger}$ & \citet{peroux-howk2020} \\
2.27 & (2, 2.5) & 1.13 & (0.85, 1.40) & \citet{zafar2013} & DLAs$^{\dagger}$ & \citet{peroux-howk2020} \\
2.73 & (2.5, 3) & 1.21 & (0.94, 1.48) & \citet{zafar2013} & DLAs$^{\dagger}$ & \citet{peroux-howk2020} \\
3.25 & (3, 3.5) & 1.26 & (0.99, 1.53) & \citet{zafar2013} & DLAs$^{\dagger}$ & \citet{peroux-howk2020} \\
3.77 & (3.5, 4) & 0.95 & (0.75, 1.15) & \citet{zafar2013} & DLAs$^{\dagger}$ & \citet{peroux-howk2020} \\
4.2 & (4, 5) & 0.96 & (0.73, 1.20) & \citet{zafar2013} & DLAs$^{\dagger}$ & \citet{peroux-howk2020} \\
2.14 & (1.55, 2.73) & 1.97 & (1.17, 3.15) & \citet{sanchez-ramirez2016} & DLAs & \citet{walter2020} \\
2.97 & (2.73, 3.21) & 0.73 & (0.38, 1.31) & \citet{sanchez-ramirez2016} & DLAs & \citet{walter2020} \\
2.15 & (2, 2.3) & 1.31 & (1.25, 1.38) & \citet{noterdaeme2012} & DLAs & \citet{walter2020} \\
2.45 & (2.3, 2.6) & 1.15 & (1.10, 1.21) & \citet{noterdaeme2012} & DLAs & \citet{walter2020} \\
2.75 & (2.6, 2.9) & 1.38 & (1.32, 1.45) & \citet{noterdaeme2012} & DLAs & \citet{walter2020} \\
3.05 & (2.9, 3.2) & 1.46 & (1.36, 1.56) & \citet{noterdaeme2012} & DLAs & \citet{walter2020} \\
3.35 & (3.2, 3.5) & 1.69 & (1.51, 1.86) & \citet{noterdaeme2012} & DLAs & \citet{walter2020} \\
2.31 & (2.2, 2.4) & 0.68 & (0.56, 0.80) & \citet{prochaska-wolfe2009} & DLAs & \citet{walter2020} \\
2.57 & (2.4, 2.7) & 0.92 & (0.82, 1.02) & \citet{prochaska-wolfe2009} & DLAs & \citet{walter2020} \\
2.86 & (2.7, 3.0) & 0.92 & (0.83, 1.02) & \citet{prochaska-wolfe2009} & DLAs & \citet{walter2020} \\
3.22 & (3, 3.5) & 1.29 & (1.18, 1.40) & \citet{prochaska-wolfe2009} & DLAs & \citet{walter2020} \\
2.35 & (2, 2.7) & 1.34 & (1.06, 1.65) & \citet{peroux2003} & DLAs & \citet{walter2020} \\
3.1 & (2.7, 3.5) & 1.23 & (0.95, 1.55) & \citet{peroux2003} & DLAs & \citet{walter2020} \\
3.9 & (3.5, 4.85) & 0.76 & (0.51, 1.03) & \citet{peroux2003} & DLAs & \citet{walter2020} \\
2.44 & (2.23, 2.6) & 1.03 & (0.94, 1.12) & \citet{noterdaeme2009} & DLAs & \citet{peroux-howk2020} \\
2.74 & (2.6, 2.88) & 1.06 & (0.98, 1.15) & \citet{noterdaeme2009} & DLAs & \citet{peroux-howk2020} \\
3.02 & (2.88, 3.2) & 1.29 & (1.19, 1.39) & \citet{noterdaeme2009} & DLAs & \citet{peroux-howk2020} \\
3.17 & (2.55, 3.4) & 1.37 & (0.90, 1.98) & \citet{guimaraes2009} & DLAs & \citet{walter2020} \\
3.62 & (3.4, 3.83) & 1.21 & (0.79, 1.68) & \citet{guimaraes2009} & DLAs & \citet{walter2020} \\
4.05 & (3.83, 5.03) & 1.07 & (0.68, 1.55) & \citet{guimaraes2009} & DLAs & \citet{walter2020} \\
3.82 & (3.5, 4.3) & 1.12 & (0.91, 1.33) & \citet{songaila-cowie2010} & DLAs & \citet{crighton2015} \\
4.45 & (4.3, 5.1) & 1.05 & (0.55, 1.76) & \citet{songaila-cowie2010} & DLAs & \citet{crighton2015} \\
4.01 & (3.56, 4.45) & 1.73 & (1.36, 2.11) & \citet{crighton2015} & DLAs & \citet{walter2020} \\
4.88 & (4.45, 5.31) & 1.44 & (1.19, 1.73) & \citet{crighton2015} & DLAs & \citet{walter2020} \\
4.99 & (4.5, 5.57) & 0.56 & (0.31, 0.82) & This work & DLAs$^{\dagger}$ & This work \\
\enddata
\tablecomments{$^{*}$This upper limit corresponds to $3\sigma$ (\citealt{kanekar2016}).}
\tablecomments{$^{\dagger}$These studies include direct measurements of the contribution from subDLAs.}
\end{deluxetable*}

\newpage

\bibliographystyle{aasjournal}
\bibliography{refs.bib}

\end{document}